%% file: main.tex
\documentclass[sigconf, nonacm, screen]{acmart}

\usepackage{amsmath}
\usepackage[noend]{algpseudocode}
\usepackage{float}
\usepackage{subcaption}
\usepackage{tabularx}
\usepackage{booktabs}
\usepackage{algorithm2e}
\usepackage{multirow}
\usepackage{booktabs}
\usepackage{threeparttable} 
\DeclareMathOperator*{\argmax}{arg\,max}
\DeclareMathOperator*{\argmin}{arg\,min}

\AtBeginDocument{%
  \providecommand\BibTeX{{%
    \normalfont B\kern-0.5em{\scshape i\kern-0.25em b}\kern-0.8em\TeX}}}

\setcopyright{acmcopyright}
\copyrightyear{2024}
\acmYear{2024}
\acmDOI{XXXXXXX.XXXXXXX}

\acmConference[SIGMOD '24]{ACM Management of Data}{June 09--15,
  2024}{Santiago, Chile}
\acmPrice{15.00}
\acmISBN{978-1-4503-XXXX-X/18/06}

\begin{document}
\pdfoutput=1

\title[ACORN: Search Over Vector Embeddings and Structured Data]{ACORN: Performant and Predicate-Agnostic Search Over
Vector Embeddings and Structured Data}


\author{Liana Patel}
\affiliation{%
  \institution{Stanford University}
  \city{Stanford}
  \country{USA}
}
\email{lianapat@stanford.edu}

\author{Peter Kraft}
\affiliation{%
  \institution{DBOS, Inc.}
  \country{USA}
}
\email{peter.kraft@dbos.dev}

\author{Carlos Guestrin}
\affiliation{%
  \institution{Stanford University}
  \city{Stanford}
  \country{USA}
}
\email{guestrin@stanford.edu}

\author{Matei Zaharia}
\affiliation{%
  \institution{UC Berkeley}
  \city{Berkeley}
  \country{USA}
}
\email{matei@berkeley.edu}

\renewcommand{\shortauthors}{}

\newif\ifcomments
\ifcomments
    \providecommand{\pb}[1]{{\color{purple}{/* peter: #1 */}}}

\else
    \providecommand{\pb}[1]{}
\fi

\newif\ifrev
\ifrev
    \providecommand{\rev}[1]{{\color{blue}{#1}}}

\else
    \providecommand{\rev}[1]{{\color{black}{#1}}}

\fi

\input{Sections/Abstract}





\maketitle

\input{Sections/Introduction}

\input{Sections/Background}

\input{Sections/Preliminaries}

\input{Sections/Algorithms}
\input{Sections/Discussion}

\input{Sections/Evaluation}

\input{Sections/RelatedWork}

\input{Sections/Conclusion}

\begin{acks}
The authors would like to thank Peter Bailis for his valuable feedback on this work.

This research was supported in part by affiliate members and other supporters of the Stanford DAWN project, including Meta, Google, and VMware, as well as Cisco, SAP, and a Sloan Fellowship. Any opinions, findings, and conclusions or recommendations expressed in this material are those of the authors and do not necessarily reflect the views of the sponsors.
\end{acks}

\bibliographystyle{ACM-Reference-Format}
\bibliography{citations}

\end{document}

%% file: Sections/Abstract.tex
\begin{abstract}
    Applications increasingly leverage mixed-modality data, and must jointly search over \emph{vector data}, such as embedded images, text and video, as well as \emph{structured data}, such as attributes and keywords. Proposed methods for this \emph{hybrid search} setting either suffer from poor performance or support a severely restricted set of search predicates (\emph{e.g.,} only small sets of equality predicates), making them impractical for many applications. To address this, we present ACORN, an approach for performant and predicate-agnostic hybrid search. 
    ACORN builds on Hierarchical Navigable Small Worlds (HNSW), a state-of-the-art graph-based approximate nearest neighbor index, and can be implemented efficiently by extending existing HNSW libraries. ACORN introduces the idea of \emph{predicate subgraph traversal} to emulate a theoretically ideal, but impractical, hybrid search strategy.
    ACORN's predicate-agnostic construction algorithm is designed to enable this effective search strategy, while supporting a wide array of predicate sets and query semantics.
     We systematically evaluate ACORN on both prior benchmark datasets, with simple, low-cardinality predicate sets, and complex multi-modal datasets not supported by prior methods.
    We show that ACORN achieves state-of-the-art performance on all datasets, outperforming prior methods with 2--1,000$\times$ higher throughput at a fixed recall. 
\end{abstract}

%% file: Sections/Introduction.tex
\section{Introduction}

Due to the representation strength of modern deep learning models, vector embeddings have become a powerful first-class datatype for wide-ranging applications that use retrieval-augmented generation
\cite{llamaindex,langchain} 
or similarity-based search \cite{borisyuk_visrel_2021,liu_pre-trained_2021,du_amazon_2022}.
As a result, vector databases and indices are seeing increasing adoption in many production use cases. 
These systems provide an efficient approximate-nearest-neighbor (ANN) search interface over embedded unstructured data \emph{e.g.,} images, text, video, or user profiles.

However, many applications must jointly query \emph{both unstructured and structured data}, requiring ANN search in combination with predicate filtering. 
For example, customers on an e-commerce site can search for t-shirts similar to a reference image, while filtering on price \cite{weianalyticdbv2020}.
Similarly, researchers performing a literature review may search with both natural language queries and filters on publication date, keywords or topics \cite{rekabsaz_tripclick_2021}.
Likewise, a data scientist working on outlier detection can find misclassified images by retrieving those that look similar to a reference dog but have the label "cat" \cite{fastdup,labelbox}.

To leverage diverse data modalities, applications need data management systems that effectively support \emph{hybrid search queries}, i.e., similarity search with structured predicates.
Such systems require
(1) \textbf{query performance}, \emph{i.e.,} efficient and accurate search despite variance in workload characteristics, such as selectivity, attribute correlations, and scale,  
and (2) \textbf{expressive query semantics}: support for diverse query predicates that may not be known a priori (\emph{e.g.,} user-entered keywords, range searches, or regex matching).

Unfortunately, existing systems fall short of these goals.
Three commonly used methods are pre-filtering \cite{wangmilvus2021,weianalyticdbv2020},  post-filtering \cite{zhang_vbase_2023, noauthor_filtered_nodate,noauthorfaiss2023, weianalyticdbv2020, wangmilvus2021}, and specialized data structures for low-cardinality predicate sets \cite{wang_navigable_2022, gollapudi_filtered-diskann_2023, wu_hqann_2022, mohoney_high-throughput_2023}. 
\emph{Pre-filtering} first finds all records in the dataset that pass the query predicate, then performs brute force similarity-search over the filtered vector set. 
This approach scales poorly, becoming inefficient for medium to high selectivity predicates on large datasets.
Alternatively, \emph{post-filtering} first searches an ANN index, then removes results that fail the query predicate. Since the database vectors closest to the query vector may not pass the predicate, post-filtering methods must typically expand the search scope. This is often expensive, particularly for search predicates with low selectivity or low correlation to the query vector, as we show in Figure \ref{fig4:correlation_and_clustering}.
Milvus \cite{wangmilvus2021}, Weaviate \cite{noauthor_filtered_nodate}, AnalyticDB-V \cite{weianalyticdbv2020}, and FAISS-IVF \cite{noauthorfaiss2023} build systems using these two core methods, and suffer from their performance limitations.

Recognizing these limitations, recent works construct \emph{specialized indices} designed for hybrid search workloads with low-cardinality predicate sets consisting of equality predicate operators.
For example, Filtered-DiskANN \cite{gollapudi_filtered-diskann_2023} outperforms prior baselines, but restricts the cardinality of the predicate set to about 1,000 and only supports equality predicates. HQANN \cite{wu_hqann_2022} and NHQ \cite{ashenon3_nhq_2023} similarly constrain the predicate set to a small number of equality filters and in addition allow only a single structured attributes per dataset entry.
These methods are often impractical since many applications have large, or unbounded predicate sets that are unknown a priori. In general, the possible predicate set's cardinality grows exponentially with each attribute's cardinality, which itself may be large.
Thus, we instead propose a \emph{predicate-agnostic} index, which can support arbitrary and unbounded predicate sets.

In this paper, we propose \textbf{ACORN} (ANN Constraint-Optimized Retrieval Network), a novel approach for performant and predicate-agnostic hybrid search that can serve high-cardinality and unbounded predicate sets. We propose two indices: \emph{ACORN-$\gamma$}, designed for high-efficiency search, and \emph{ACORN-1}, designed for low construction overhead in resource-constrained settings. 
Both methods modify the hierarchical navigable small world (HNSW) index, a state-of-the-art graph-based index for ANN search, and are easy to implement in existing HNSW libraries.

ACORN tackles both the \emph{performance limitations} of pre- and post- filtering, as well as the \emph{semantic limitations} of specialized indices. 
ACORN proposes the idea of \emph{predicate subgraph traversal} during search.
As the name implies, the search strategy traverses the subgraph of the ACORN index induced by the set of nodes that pass the query predicate. ACORN designs the index such that these arbitrary predicate subgraphs approximate an HNSW index. 
Unlike pre- and post-filtering, this allows ACORN to provide sublinear retrieval times despite variance in \emph{correlation} between query vectors and predicates, which we find to be a major challenge for existing hybrid search systems.
ACORN also serves wide-ranging predicate sets by employing a predicate-agnostic construction that alters HNSW's algorithm to create a denser graph.
Specifically, we introduce a predicate-agnostic neighbor expansion strategy in ACORN-$\gamma$ based on target predicate selectivity thresholds, which can be estimated empirically with or without knowing the predicate set.
In conjunction, we propose a predicate-agnostic compression heuristic to efficiently manage the index space footprint while maintaining efficient search. 
We also explore the trade-off space between search performance and construction overhead, designing ACORN-1 to approximate ACORN-$\gamma$'s search performance while further reducing the time-to-index (TTI) for resource-constrained settings.

We systematically evaluate ACORN-$\gamma$ and ACORN-1 on four datasets: SIFT1M \cite{jegou_product_2011}, Paper \cite{wang_navigable_2022}, LAION \cite{schuhmann_laion-400m_2021}, and TripClick \cite{rekabsaz_tripclick_2021}.
Our evaluation includes both prior benchmark datasets, with simple, low-cardinality predicate sets, which prior specialized indices can serve, as well as more complex datasets with millions of possible predicates, which existing indices cannot to handle. 
On each, ACORN-$\gamma$ achieves state-of-the-art hybrid search performance with 2--1000$\times$ higher queries per second (QPS) at 0.9 recall compared to prior methods. 
\rev{Specifically, ACORN achieves 2-10$\times$ higher QPS on prior benchmarks, over 30$\times$ higher QPS on new benchmarks, and over 1,000x higher QPS at scale, on a 25-million-vector dataset.} We find that ACORN-1 empirically approximates ACORN-$\gamma$, attaining at most $5\times$ lower QPS at fixed recall but 9--53$\times$ lower TTI compared to ACORN-$\gamma$. 
Our detailed evaluation demonstrates the effectiveness of ACORN's predicate-subgraph traversal strategy and predicate-agnostic construction techniques.

%% file: Sections/Background.tex
\section{Background}
Existing methods for Approximate Nearest Neighbor (ANN) search can be broadly categorized as tree-based  \cite{bentley_multidimensional_1975, bernhardsson_annoy_nodate, beygelzimer_cover_2006, dasgupta_random_2008, houle_rank-based_2015, lu_vhp_2020, muja_scalable_2014, silpa-anan_optimised_2008}, hashing-based  \cite{andoni_near-optimal_2008, andoni_practical_2015, andoni_optimal_2015, gong_idec_2020, indyk_approximate_1998, jafari_mmlsh_2020, li_io_2020, liu_ei-lsh_2021, lu_r2lsh_2020, lv_intelligent_2017, park_neighbor-sensitive_2015, sundaram_streaming_2013, zheng_pm-lsh_2020, gionis_similarity_1999}, quantization-based  \cite{ge_optimized_2014, guo_accelerating_2020, jegou_product_2011, johnson_billion-scale_2017, lempitsky_inverted_2012}, and graph-based  \cite{fu_fast_2019, malkov_approximate_2014, gollapudi_filtered-diskann_2023, jayaram_subramanya_diskann_2019, malkov_efficient_2018, singh_freshdiskann_2021, zhao_song_2020}. 
In this work build on HNSW, a graph-based method that is empirically one of the best-performing on high-dimensional datasets, and we adapt it to support hybrid search.

Graph-based ANN methods have gained popularity due their state-of-the-art  performance on varied ANN benchmarks \cite{aumuller_ann-benchmarks_2020, simhadri_results_2022}. 
These methods typically perform search using a greedy routing strategy that traverses a graph index, starting from a pre-defined entry point. The index itself forms a proximity graph $G(V,E)$, such that each dataset point is represented  by a vertex and edges connect nearby points.
The index construction algorithm typically aims to approximate subgraphs of the Delaunay graph \cite{lee_two_1980}.
While the Delaunay graph guarantees convergence of a greedy routing algorithm, it is impossible to efficiently construct for arbitrary metric spaces \cite{navarro_searching_2002}.
Thus graph methods focus on more tractable approximations of Delaunay subgraphs, such as the \emph{Relative Neighbor Graph (RNG)} \cite{toussaint_relative_1980, lankford_regionalization_1969}, and the \emph{Nearest-Neighbor Graph (NNG)} \cite{dong_efficient_2011, algorithms_kgraph_2023}.

\subsection{Hierarchical Navigable Small Worlds}
\label{sec:graph-based-methods} 

\begin{figure}[t]
  \centering
  \includegraphics[width=0.30\linewidth]{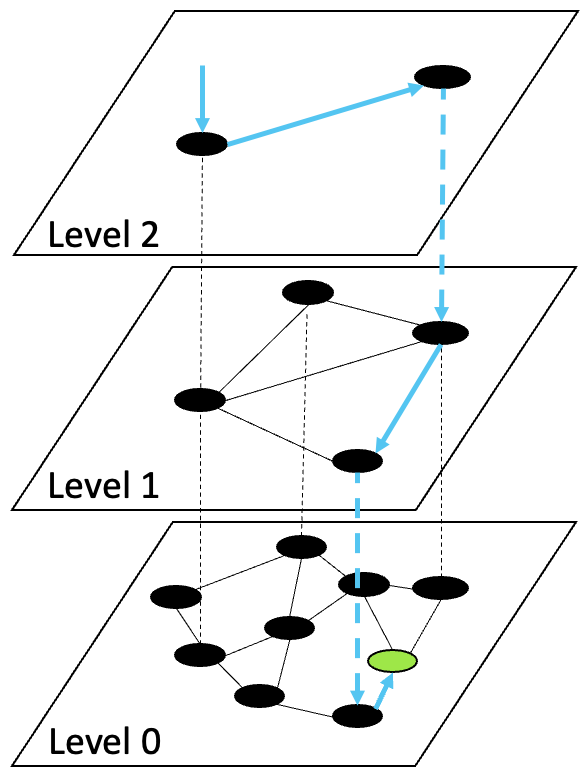}
  \vspace{-0.2cm}
  \caption{\small Schematic drawing of search over an HNSW index. The search path is shown by blue arrows, beginning on level 2 and ending on level 0 at the query point, shown in green.}
  \label{fig:hnsw_diagram}
  \vspace{-.5cm}
\end{figure}

As Figure \ref{fig:hnsw_diagram} illustrates, HNSW forms a hierarchical, multi-level graph index with bounded degree.
Below we briefly summarize the HNSW search and construction algorithm. 

\emph{\textbf{The HNSW construction algorithm}} iteratively inserts each point into the graph index, to construct a navigable graph with bounded degree, specified by parameter $M$.
For each inserted element $v$, first, a maximum layer index $l$ is stochastically chosen using an exponentially decaying probability distribution, normalized by the constant $m_L = 1/ln(M)$. The level assignment probability ensures that the expected characteristic path length increases with the level index. Intuitively, the upper-most level contains the longest-range links, which will be traversed first by the search algorithm, and the bottom-most level contains the shortest-range links, which are traversed last by the search algorithm.
The insertion procedure then proceeds in two phases. In the first phase, a greedy search is performed iteratively from the top layer, beginning at a pre-defined entrypoint down to the $(l+1)$th layer. 
At each of these levels, the greedy subroutine chooses a single node that becomes the entry-point into the next layer. 
In the second phase, the greedy search iterates over level $l$ to level 0. 
The greedy search at each level now chooses $\textit{efc}$ nodes as candidate edges. 
Of these candidates, at most $M$ are selected to become neighbors of $v$ according to an RNG-based pruning algorithm \cite{jaromczyk_relative_1992}.
At level 0, the degree bound is increased $2\times M$, which is shown to empirically improve performance.

\emph{\textbf{The HNSW search algorithm}} begins its traversal from a pre-defined entry point at the upper-most layer of the multilayer graph, illustrated in Figure \ref{fig:hnsw_diagram}.
The traversal then follows an iterative search strategy from the top level downwards. At each level a greedy search is used to choose a single node, which becomes the entry-point into the next level. Once the bottom level is reached, rather than greedily choosing a single node, the search algorithm greedily chooses $K$ nearest elements to return. We outline this process in Algorithm \ref{alg:hnsw-ann-search}. The search parameter $\textit{efs}$ provides a tradeoff between search quality and efficiency by controlling the size of the dynamic candidate list stored during the bottom level's greedy search.

\input{algs/hnsw-ann-search}

%% file: algs/hnsw-ann-search.tex
\RestyleAlgo{ruled}

\SetKwComment{Comment}{// }{} 

\begin{algorithm}[t!]
\small
\caption{HNSW-ANN-SEARCH($x_q, K, \textit{efs}$)}\label{alg:hnsw-ann-search}
\KwIn{query vector $x_q$, number of nearest neighbors to return $K$, size of dynamic candidate list $\textit{efs}$}
\KwOut{$K$ nearest elements to $x_q$}
$e \gets$ entry-point to hnsw graph\\
$W \gets \emptyset$ \Comment{set of current nearest}\
$L \gets level(e)$ \Comment{Top hnsw level}\ 
\For{$l \gets L...1$}{
    $e \gets$ SEARCH-LAYER($x_q, e, \textit{ef}=1, l$)\
  }\
$W \gets$ SEARCH-LAYER($x_q, e, \textit{ef}=\textit{efs}, l=0$)\

\Return{$K$ nearest elements from $W$ to $x_q$}\
\end{algorithm}


%% file: Sections/Preliminaries.tex
\section{Problem Definition and Challenges}

In this section we formally define the hybrid search setting and then analyze the performance challenges that existing predicate-agnostic methods, i.e., pre- and post-filtering, face. 
Our analysis leads us to explore several important workload characteristics. Specifically, we will consider predicate selectivity, the dataset size, and \emph{query correlation}, which we introduce, formally define and find to be a major challenge for post-filtering methods.

\rev{We will later leverage our understanding of existing performance challenges in Section \ref{sec:prelim_theoretical_ideal} to formulate a theoretically ideal hybrid search solution. 
Then, in Section \ref{sec:eval}, we will revisit the workload characteristics discussed in this section to rigorously evaluate ACORN's search performances.}

\subsection{Hybrid Search Definitions}\label{subsec:hybrid_search_def}
Let $D = \{e_1, e_2, ..., e_n\} = \{(x_1, a_1), (x_2, a_2), ..., (x_n, a_n)\}$ be a dataset consisting of $n$ entities, each with a vector component, $x_i \in \mathbb{R}^d$, and a structured attribute-tuple, $a_i$, associated with entity $e_i$. 
Let $X = \{x_1, x_2, ..., x_n\}$ denote the set of vectors in the dataset, and $dist(a,b)$ is the metric distance between any two points. Let $A = \{a_1, a_2, ..., a_n\}$ be the set of structured attributes in the dataset. We will denote $X_p \subseteq X$ as the subset of vectors corresponding to entities in the dataset that pass a given predicate $p$. We refer to the \emph{selectivity} $(s)$ of predicate $p$ as the fraction of entities from $D$ that satisfy the predicate, where $0 \leq s \leq 1$.

We consider the \emph{\textbf{hybrid search problem}}, described as follows. 
Given a dataset $D$, target $K$, and query $q = (x_q, p_q)$, where $x_q \in \mathbb{R}^d$, and $p_q$ is a predicate, retrieve $x_q$'s $K$ nearest neighbors that pass the predicate $p_q$.  
We will specifically focus on the problem of \emph{approximate} nearest neighbor search w.r.t $x_q$. 
Here, our goal is to maximize both search accuracy and search efficiency. We will measure accuracy by $\textit{recall}@K = \frac{G \cap R}{K}$, where $G$ is the ground truth set of $K$ nearest neighbors to $x_q$ that satisfy $p_q$, and $R$ is the retrieved set.

\subsection{Search Performance of Baseline Methods}\label{subsec:prelim_hybrid_search_perf} 
We now analyze the search complexity of two predominant baseline methods, pre- and post-filtering. We will consider how varied workload characteristics impact the search behavior of these methods.
Through our analysis, we will make the standard assumption that distance computations dominate search performance.
We note that HNSW's unfiltered search complexity is $O(\log(n) + K)$.

\emph{\textbf{Pre-filtering}} linearly scans $X_p$, computing distances over each point that passes the search predicate. This yields a hybrid search complexity of $O(|X_p|) = O(sn + K)$.
While pre-filtering always achieves perfect recall, its search complexity scales poorly for large dataset sizes or selectivities, growing linearly in either variable.

\emph{\textbf{Post-filtering}}, by contrast, performs ANN-search over $X$ to find the closet query vector to $x_q$, then expands the search scope to find $K$ vectors that pass the query predicate, $p$. 
Intuitively, search performance varies depending on \emph{correlation} between the query vector and the vectors in $X_p$.
When the vectors of $X_p$ are close to the query vector, post-filtering over HNSW has a search complexity of $O(\log(n) + K)$. 
If the vectors in $X_p$ are uniformly distributed within $X$, then post-filtering's expected search complexity is $O(\log(n) + K/s)$. 
However, vectors of $X_p$ may be far away from the query vector, leading to a worst case of $O(n)$ search performance.

We see that the search performance of either baseline \emph{is not robust} to variations in selectivity, dataset size, and query correlation. 
We empirically verify these limitations in section \ref{sec:eval} (figures \ref{fig:tripclick}, \ref{fig2:laion1m_correlation}).

\subsubsection{Formalizing Query Correlation}\label{subsec:prelim_workload_characteristics} 
\begin{figure}[t]
  \centering
  \includegraphics[width=0.7\linewidth]{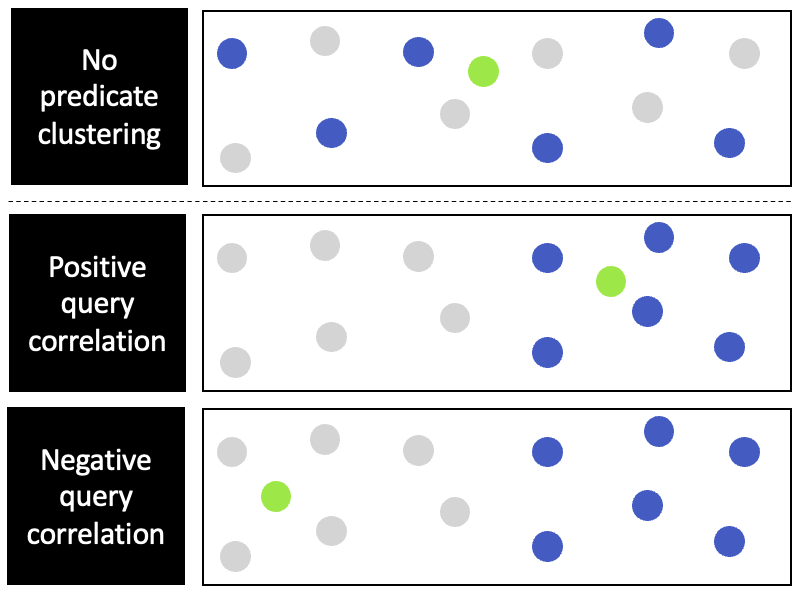}
  \vspace{-0.2cm}
  \caption{\small Schematic drawing of a dataset with no predicate clustering (top), a dataset with predicate clustering and positive query correlation (middle), and a dataset with predicate clustering and negative query correlation (bottom). Dark blue circles show points that pass the predicate, and light gray circles show points that fail the predicate. The query vectors are shown in green.}
  \label{fig4:correlation_and_clustering}
  \vspace{-.5cm}
\end{figure}

We will now formalize the notion of \emph{query correlation}, which we find is key challenge for post-filtering-based systems. As Figure \ref{fig4:correlation_and_clustering} shows, query correlation occurs when the vectors of $X_p$ are non-uniformly distributed in $X$ and instead cluster together relative to the vectors in X. We refer to this phenomenon as \emph{predicate clustering}. When predicate clustering occurs, a query vector may be either close or far away from the predicate cluster containing its search targets, inducing query correlation.

\emph{\textbf{Definition: Query Correlation.}}
\rev{
We will consider the query-to-target distances for the \emph{given dataset} compared to the expected query-to-target distances for a \emph{hypothetical dataset}, under which no clustering is present.
Formally, we define the \emph{query correlation} of the hybrid search workload $Q$ over dataset $D$ as:
$$C(D, Q) = \mathbb{E}_{(x_i, p_i) \in Q}\bigl[\mathbb{E}_{R_i}[g(x_i, R_i)] - g(x_i, X_{p_i})\bigr]$$

We let $R_i$ be a random set variable of $|X_{p_i}|$ vectors drawn \emph{uniformly} from $X$, defined for each hybrid query $(x_i, p_i) \in Q$.
We define $g(x, S) = \min_{y\in S} dist(x, y)$  to be the function mapping the query vector $x$ to the minimum distance of neighbors from the given vector set $S \subseteq R^d$.
Note that $g(x_i, X_{p_i}$) is the ground-truth hybrid-search target of the query $(x_i, p_i)$.

If, on average, query vectors are closer to their targets in $X_{p_i}$, the true dataset of hybrid search targets, than in $R_i$, the no-clustering dataset, then the workload has \emph{positive query correlation}. If the reverse is true, the workload has \emph{negative query correlation}.  
We may also consider nearest-neighbor distance rather than the metric distance in the above definition. 
We also note that we can easily extend this definition to consider $K$ targets of the hybrid search, rather than one, by summing distances over the $K$ search targets. 
}


\section{Theoretical Ideal Hybrid Search Performance with HNSW}\label{sec:prelim_theoretical_ideal}
For a given hybrid search query, we define the theoretically ideal search performance using HNSW data structures as the performance attainable if we knew the search predicate $p_q$ during construction.
In this case, we could construct an HNSW index over $X_p$. 
We call this the \emph{\textbf{oracle partition index}} for that query.
The complexity of searching this index is $Os(\log sn + K)$. 
Notably, the search performance of the oracle partition index outperforms both pre- and post-filtering across variations in predicate selectivity, data size, and query correlation. While pre-filtering's search scales in $|X_p|$, search over the oracle partition scales \emph{sublinearly} in $|X_p|$. The oracle partition is also robust under variations in query correlation: it does not require the search scope expansion used in post-filtering. 

Despite its ideal search performance, the oracle partition index requires us to know all search predicates in advance and to create a full HNSW index per predicate. In practice, the oracle partition index is not possible to construct because query predicate sets are often unknown during construction and have high or unbounded cardinality. Building an HNSW per predicate would require prohibitive amounts of space and time. Thus, in this work, we will instead approximate search over the oracle partition index for a particular query, without ever explicitly constructing this index.

%% file: Sections/Algorithms.tex
\section{ACORN Overview}

\input{algs/formatting}

\begin{figure}[t!]
    \centering
    \vspace{-.5cm}    \includegraphics[width=0.3\linewidth]{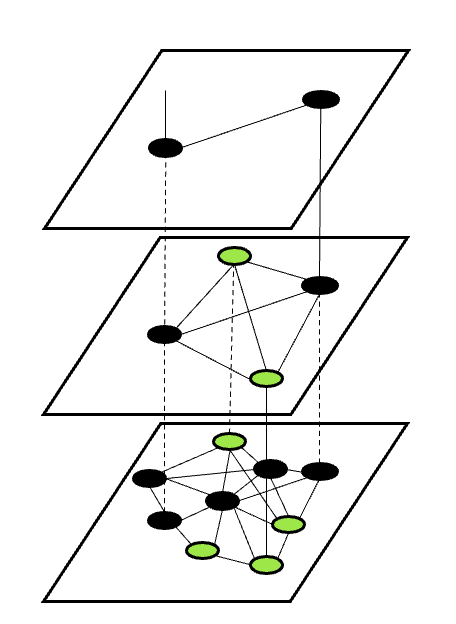}
    \vspace{-.3cm}
    \caption{\small An illustration of the predicate subgraph, shown by the green nodes. ACORN searches over the predicate subraph to emulate search over an oracle partition index.}
    \label{fig:predicate subgraph}
    \vspace{-.5cm}
\end{figure}

\begin{figure*}[hbt!]
  \centering
  \includegraphics[width=0.7\linewidth]{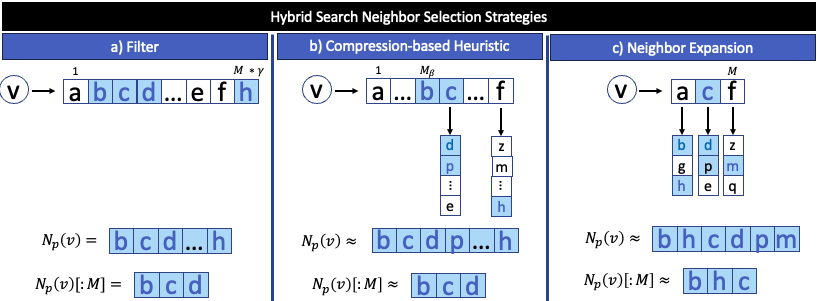}
  \caption{\rev{\small Diagram of ACORN's neighbor selection strategies. Blue nodes represent neighbors that pass the query predicate. Sub-figure (a) shows the simple predicate-based filter applied to uncompressed edge lists of size $M\cdot \gamma$, followed by truncation to size $M=3$. Sub-figure (b) shows the compression-based heuristic. Sub-figure (c) shows the neighbor expansion strategy used in ACORN-1.}}
  \label{fig:search_diagram}
  \vspace{-.3cm}
\end{figure*}

We now describe ACORN, a predicate-agnostic approach for state-of-the-art hybrid search. 
We propose two variants, which we refer to as \textbf{ACORN-$\gamma$} (\ref{subsec:acorn-gamma-search}, \ref{subsec:acorn-gamma-construction})
and \textbf{ACORN-1} (\ref{subsec:acorn-1}). 
We design ACORN-$\gamma$ to achieve efficient search performance, and we design ACORN-1 to approximate ACORN-$\gamma$'s search performance while further reducing the algorithm's time to index (TTI) and space footprint for resource-constrained settings.

ACORN's core idea is to search over the index's \emph{predicate subgraph}, i.e., the subgraph induced by $X_p$ for a given search predicate $p$, as shown in Figure \ref{fig:predicate subgraph}. 
We modify the HNSW construction algorithms so that arbitrary predicate subgraphs emulate an HNSW oracle partition index without the need to explicitly construct one. 
ACORN-$\gamma$ achieves this by constructing a denser version of HNSW, which we parameterize by a neighbor list expansion factor, $\gamma$, a compression factor, $M_\beta$, and the HNSW parameters, $\textit{efc}$ and $M$.
Then by adding a filter step during search to ignore neighbors that fail the predicate, we find ACORN-$\gamma$'s search can efficiently navigate to and traverse over the predicate subgraph, even under variations in query correlation. Meanwhile, ACORN-$1$ expands neighbor lists \emph{during search} rather than during construction to \emph{approximate} ACORN-$\gamma$'s dense graph structure without building it.\footnote{For highly selective queries where even ACORN's predicate subgraph would be disconnected within the larger ACORN graph, ACORN falls back to pre-filtering, which is effective for such queries. ACORN is configured with a minimum selectivity, $s_{min}$, under which it should use pre-filtering when a query is estimated to be more selective than $s_{min}$. We describe how to configure $\gamma$ based on $s_{min}$ in Section~\ref{subsec:acorn-gamma-construction}.}

Overall, ACORN prescribes a simple and general framework for performant hybrid search based on the idea of \emph{predicate-subgraph traversal}. The core techniques we propose are \emph{predicate-agnostic} neighbor-list expansions and pruning during construction in combination with predicate-based filtering during search. 
While this framework can be applied to a variety of graph-based ANN indices, in this work we focus on HNSW due their state-of-the-art performance and widespread use.

\input{tables/notation}

\subsection{ACORN-$\gamma$ Search Algorithm}\label{subsec:acorn-gamma-search}
Algorithm \ref{alg:acorn-search-layer} outlines the greedy search algorithm ACORN uses at each level, beginning from the top level at a pre-defined entry-point. The main difference between ACORN's search algorithm and that of HNSW is how neighbor look-ups (line 9) are performed at each visited node, $c$. While HNSW simply checks the neighbor list, $N^l(c)$, ACORN performs additional steps to recover an appropriate neighborhood for the given search predicates.

\input{algs/ACORN-search-layer}
Specifically, ACORN-$\gamma$ uses two neighbor look-up strategies, a simple filter method, shown in Figure \ref{fig:search_diagram}(a), and a compression-based heuristic, shown in Figure \ref{fig:search_diagram}(b), which is compatible with the compression strategy we optionally apply during construction, detailed in Section \ref{subsec:acorn-gamma-construction}.
For each visited node, $v$, the filter-based neighbor look-ups simply scan the neighbor list $N^l(v)$ to find the sub-list of neighbors that pass the predicate, $N_p^l(v)$. If $N_p^l(v)$ contains more than $M$ nodes, we take the first $M$ and return this as $v$'s neighborhood. The compression-based neighbor look-ups instead partially expand the neighbor set $N^l(v)$ to include a subset of $v$'s two-hop neighbors, before performing filtering and truncation. This procedure entails two phases. The first phase iterates through the first $M_\beta$ nodes of $N^l(v)$, simply filtering as in the previous strategy. The second phase iterates over the remainder of the neighbor list, expanding the search neighborhood to include neighbors of neighbors, before again filtering according to the query predicate. $M_\beta$ is a construction parameter which we will discuss in the next section.

\subsection{ACORN-$\gamma$ Construction Algorithm}\label{subsec:acorn-gamma-construction}

\begin{figure}[!t]
  \centering
  \vspace{-.4cm}
  \includegraphics[width=1.0\linewidth]{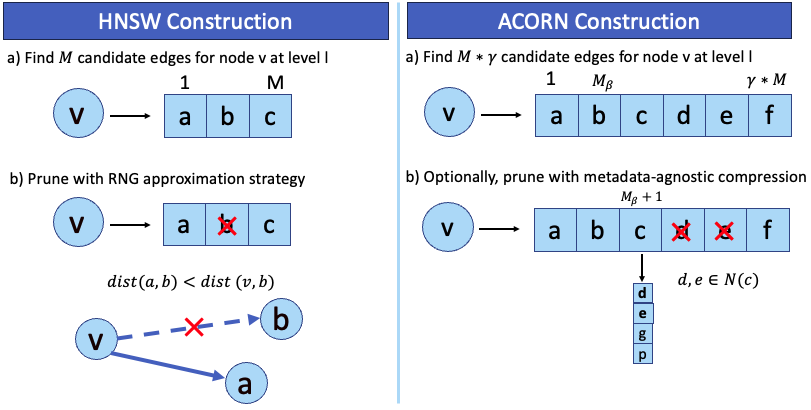}
  \vspace{-.5cm}
  \caption{\small A comparison of HNSW and ACORN-$\gamma$'s strategies for (a) selecting candidate edges, shown for $M$=3, and (b) pruning candidate edges for each inserted node $v$, shown for $M$=3, $M_{\beta}$=2, $\gamma$=2.}
  \label{fig:construction_diagram}
   \vspace{-.3cm}
\end{figure}

We construct the ACORN-$\gamma$ index by applying two core modifications to the HNSW indexing algorithm: first, we expand each node's neighbor list, and \rev{then we apply a novel predicate-agnostic pruning method to compress the index.} Both of these steps are summarized in Figure \ref{fig:construction_diagram}.

\emph{\textbf{Neighbor List Expansion.}} 
While HNSW collects $M$ approximate nearest neighbors as candidate edges for each node in the index, ACORN collects $M\cdot \gamma$ approximate nearest neighbors as candidate edges per node. 
\rev{To find these candidates during construction, ACORN uses a metadata-agnostic search over its graph index.
Specifically, the neighbor lookup strategy at each node, $v$, on level $l$, simply accesses the neighbor list $N^l(v)$ and returns the first $M$ nodes.
Note that although each node contains up to $M\cdot \gamma$ neighbors, we assume by construction that $M$ neighbors per node are sufficient for maintaining navigability of the graph index. 
Thus, considering truncated neighbor lists while traversing the graph allows us to avoid unnecessary distance computations and TTI slowdowns.}

One simple choice for $\gamma$ is $\frac{1}{s_{min}}$, where $s_{min}$ is the minimum predicate selectivity we plan to serve before resorting to pre-filtering. 
\rev{As we discuss in Section \ref{sec:discussion}, ACORN's indexing time and space footprint increase proportionally to $\gamma$. Meanwhile, pre-filtering becomes a competitive baseline at low predicate selectivity values, as we show in Figure \ref{tripclick_1p}. Thus, ACORN is able to balance construction and search efficiency by using pre-filtering as a fall-back for queries with low-selectivity predicates. }
This leads to a simple \emph{cost-based model} during search: if the estimated predicate selectivity of a given query is greater than $1/\gamma$, search the ACORN-$\gamma$ index, otherwise pre-filter. 
\rev{We note that leveraging pre-filtering in this way may degrade search efficiency, but not result quality, when errors occur in selectivity estimates. If a query's true predicate selectivity is above $1/\gamma$, but the estimate is below, the system will mistakenly pre-filter, obtaining perfect recall at possibly lower QPS than if the ACORN index was instead searched. If the reverse is true, the system will mistakenly search the ACORN index, whereas pre-filtering would have offered similar QPS and perfect recall.}

\emph{\textbf{Compression.}}
A key challenge with ACORN-$\gamma$'s neighbor expansion step is that it increases index size and TTI. The increased index size poses a significant issue particularly for memory-resident graph indices, like HNSW. To address this, we introduce a \emph{predicate-agnostic pruning} technique. While we could apply compression to the full index, as discussed in Section \ref{subsec:discussion_indexsize}, we specifically target the bottom level's neighbor lists since they contribute most significantly to the indexing overhead. This follows from the exponentially decaying level assignment probability ACORN uses.

The core idea of the pruning procedure is to precisely retain each node's nearby neighbors in the index, while approximating farther away neighbor during search. 
\rev{We use the \emph{tunable compression parameter}, $M_\beta$, where $0\leq M_{\beta}, \leq M \cdot \gamma$. During construction, ACORN chooses each node's final neighbor list by automatically retaining the nearest $M_\beta$ candidate edges and aggressively pruning the remaining candidates.}
During search we can recover the first $M_\beta$ neighbors of each node $v$ directly from the neighbor list $N^l(v)$, and approximate remaining neighbors by looking at 2-hop neighbors during search, as we described in Section \ref{subsec:acorn-gamma-search}.

Figure \ref{fig:construction_diagram} outlines this pruning procedure applied to node $v$'s candidate neighbor list. The algorithm iterates over the ordered candidate edge list and keeps the first $M_\beta$ candidates. \rev{Over the remaining sub-list of candidates, the algorithm applies the following pruning procedure at each node. Let H be the dynamic set of $v$'s chosen two-hop neighbors, initialized to $\emptyset$. We prune candidate $c$ if it is contained in $H$; otherwise, we keep $c$ and add all of its neighbors to $H$. The pruning procedure stops after iterating over all candidates, or if $|H|$ plus the number of chosen edges exceeds $M\cdot \gamma$. The pruned and ordered neighbor list is then stored in the ACORN index and H is discarded.}

\rev{We highlight that the neighbor expansion during search, described in Section 5.1, can recover pruned neighbors regardless of the query predicate. It follows from ACORN's pruning rule that any node $x$ that was pruned from some node $v$'s neighbor list, $N^l(v)$, must be in the neighbor list $N^l(y)$ such that $y$ is a neighbor of $v$ with index greater than $M_\beta$. During search, the neighbor lookup at $v$ on level $l$ will perform a neighbor-list expansion for all neighbors with an index greater than $M_\beta$, thus checking $N^l(y)$ and finding $x$.}

We now briefly describe why HNSW's pruning, a \emph{metadata-blind} mechanism, is insufficient for hybrid search. Consider the simple scenario shown in Figure $\ref{fig:construction_diagram}$. For a node, $v$, inserted into the HNSW index at an arbitrary level $l$, the algorithm generates candidates neighbors $a, b$  and $c$. HNSW's pruning rule iterates over $v$'s candidate neighbor list in order of nearest to farthest neighbors. Node $b$ is pruned since there exists a neighbor $a$ such that  $b$ is closer to $a$ than to $v$. This RNG-approximation strategy corresponds to pruning the longest edge of the triangle formed by a triplet $v, a, b$. In this case, we can prune the edge $v - b$ and expect a search path to traverse from $v$ to $b$ via $a$. The problem with this technique arises when we consider the hybrid search setting for an arbitrary predicate. Say $v$ and $b$ pass a given query predicate $p_q$, but $a$ does not. Then $v, b, a$ do not form a triangle in the predicate subgraph, and we cannot expect to find the path from $v$ to $b$ through $a$. As a result, HNSW's pruning mechanism will falsely prune edge $v-b$. If we had complete knowledge of all possible query predicates, we could ensure that we only prune edges of triangles such that all three vertices always exist in the same subset of possible predicate subgraphs. FilteredDiskANN \cite{gollapudi_filtered-diskann_2023} takes this approach by restricting the set of possible query predicates. However, for arbitrary query predicates, ensuring this property holds becomes intractable.

\subsection{ACORN-1}\label{subsec:acorn-1}
We now describe ACORN-1, an alternative approach which aims to approximate ACORN-$\gamma$'s search performance, while further minimizing index size and TTI. ACORN-1 achieves this by performing the neighbor expansion step solely during search, rather than during construction, as ACORN-$\gamma$ does. 
ACORN-1's construction corresponds to the original HNSW index without pruning. This construction corresponds to ACORN-$\gamma$'s construction algorithm, with fixed parameters $\gamma = 1$ and $M_\beta = M$.

ACORN-1's main difference from ACORN-$\gamma$ during search, is its neighbor lookup strategy. Specifically, at each visited node, $v$, during greedy search, ACORN-1 uses a full neighbor list expansion to consider all one-hop and two-hop neighbors of $v$, before applying the predicate filter and truncating the resulting neighbor list to size $M$. Figure \ref{fig:search_diagram}(c) outlines this procedure.

%% file: algs/formatting.tex
\DontPrintSemicolon

\SetKwIF{If}{ElseIf}{Else}{if}{}{else if}{else}{endif}
\SetKwFor{For}{for}{}{}
\SetKwFor{ForEach}{for each}{}{}
\LinesNumbered

%% file: tables/notation.tex
\begin{small}

\begin{table}[!tb]
\rev{
  \centering
  \caption{\small Summary of Notation}
  \vspace{-.3cm}
  \begin{tabular}{cc}
    \toprule
    \textbf{Symbol} & \textbf{Description} \\
    \midrule
    $\gamma$ & neighbor expansion factor for ACORN index \\
    $M_\beta$ & compression parameter for ACORN index \\
    $\textit{ef}$ & size of dynamic candidate list in ACORN greedy search \\
    $M$ & degree bound for traversed nodes during ACORN search \\
    $m_L = 1/\ln M$ & level normalization constant for ACORN index \\
    $e$ & entry-point to ACORN index \\
    $e_p$ & entry-point to ACORN's predicate $p$'s subgraph \\
    $l(v)$ & maximum level index of node $v$ in ACORN index \\
    $N^l(v)$ & neighbor list of node $v$ at level $l$ \\
    $N^l_p(v)$ & filtered neighbors of node $v$ at level $l$ under predicate $p$ \\
    $X_p$ & vector dataset that passes predicate $p$ \\
    $s$ & selectivity \\
    $n$ & size of dataset \\
    \bottomrule
  \end{tabular}
  \vspace{-.4cm}
}
\end{table}

\end{small}

%% file: algs/ACORN-search-layer.tex
\RestyleAlgo{ruled}


\SetKwComment{Comment}{// }{} 

\begin{algorithm}[!t]
\small
\caption{ACORN-SEARCH-LAYER($x_q, p_q, e, ef, l$)}\label{alg:acorn-search-layer}
\KwIn{query vector $x_q$, query predicate $p_q$, entry-point $e$, number of nearest neighbors to return $ef$,  level to search $l$}
\KwOut{$ef$ nearest elements to $x_q$}
$T \gets e$ \Comment{visited set}
$C \gets e$ \Comment{candidate set}
$W \gets e$ \Comment{dynamic list of found nearest neighbors}

\While{$|C| > 0$}{
    $c \gets$ extract $\argmin_{x\in C} \lVert x_q - x\rVert$ \\
    $f \gets$ get $\argmax_{x\in W} \lVert x_q - x\rVert$ \\
    \uIf{$dist(c, x_q) > dist(f, x_q)$ and $|W| \geq efc$} {
        break
    }
    $neighborhood \gets$ GET-NEIGHBORS($c, l, p_q$)\\
    \ForEach{$v\in$ neighborhood$[1{:}M]$}{
        \uIf{$v\notin T$}{
            $T\gets T\cup v$\\
            $f \gets \argmax_{x\in W} \lVert x_q - x\rVert$ \\
            \uIf{$dist(v, x_q) < dist(f, x_q)$ or $|W| < ef$}{
                $C \gets C\cup v$\\
                $W \gets W \cup v$\\
                \uIf{$|W| > ef$} {
                    remove furthest element from $W$ to $x_q$
                }   
            }
        }
    }\

}
\Return{$W$}\
\end{algorithm}


%% file: Sections/Discussion.tex
\section{Discussion}\label{sec:discussion}

In this section we analyze the ACORN index's space complexity, construction complexity and search performance. We focus our attention on ACORN-$\gamma$, since ACORN-1's index construction represents a special case of ACORN-$\gamma$ for fixed parameters ($\gamma=1$, $M_{\beta}=M$), and we empirically show that ACORN-1 search approximates ACORN-$\gamma$ in Section \ref{sec:eval}. \rev{We note that our analysis in Sections \ref{subsec:discussion_construction} and \ref{subsec:discussion_search} considers the complexity scaling of the search procedure under the assumption that we build the exact Delaunay graphs rather than approximate ones.}

\subsection{Index Size}\label{subsec:discussion_indexsize}

The average memory consumption per node of the ACORN-$\gamma$ index is $O(M_\beta + M + m_L\cdot M\cdot \gamma)$, assuming the number of bytes per edge is constant.
For comparison, average memory consumption per node for the HNSW index scales $O(M + m_L\cdot M)$. Overall, ACORN-$\gamma$ increases the bottom-level's memory consumption by $O(M_\beta)$ per node, and increases the higher levels memory consumption by a factor of $\gamma$ per node.  

To understand ACORN's memory consumption we evaluate the average number of neighbors stored per node.
At level 0, compression is applied to the candidate edge lists of size $M*\gamma$ resulting in neighbor sets of length $M_\beta$ plus a compressed set which scales $O(M)$. We show this empirically in figure \ref{fig:pruning_comparison}. On higher levels, nodes have at most $M*\gamma$ edges. 
We multiply this by the average number of levels that an element is added to, given by $\mathbb{E}[l + 1] = E[-\ln(\text{unif}(0,1)) * m_L] = m_L + 1$. 

While we specifically target compression to level 0 in this work, because it uses the most space, compression could be applied to more levels in bottom-up order to further reduce the index size for large datasets. Denoting $nc$ as the chosen number of compressed levels, the average memory consumption per node in this generalized case is $O(nc(M_{\beta} + M) + (m_L - nc)(M\cdot \gamma)$.

\subsection{Construction Complexity}\label{subsec:discussion_construction}
For fixed parameters $M, M_\beta$ and $\textit{efc}$, ACORN-$\gamma$'s overall expected construction complexity scales $O(n\cdot \gamma\cdot \log(n)\cdot \log(\gamma))$. Compared to HNSW, which has $O(n\cdot \log(n))$ expected construction complexity, ACORN-$\gamma$ increases TTI by a factor of $\gamma\cdot \log(\gamma))$ due to the expanded edge lists it generates.

We now describe ACORN's construction complexity in detail  by decomposing it into the following three factors \textbf{(i)} the number of nodes in the dataset, given by $n$ \textbf{(ii)} the expected number of levels searched to insert each node into the index, and \textbf{(iii)} the expected complexity of searching on each level. By design, ACORN's expected maximum level index scales $O(\log n)$ according to its level-assignment probability, which is the same as HNSW. This provides our bound on (ii). 

Turning our attention to (iii), we will first consider the length of the search path and then consider the computation cost incurred at each visited node.
For the HNSW level probability assignment, it is known that the expected greedy search path length is bounded by a constant $S=\frac{1}{1- \exp(-m_L)}$ \cite{malkov_efficient_2018}. 
We can bound ACORN's expected search path length by $O(\gamma)$ since the path reaches a greedy minima in a constant number of steps and proceeds to expand the search scope by at most $M\cdot \gamma$ nodes to collect up to $M\cdot \gamma$ candidate neighbors during construction. 

The computation complexity at each visited node along the search path is $O(\log(\gamma))$, seen as follows. For each node visited, we first check its neighbor list to find at most $M$ un-visited nodes, on which we perform distance computations in $O(M\cdot d)$ time. Then, we update the sorted lists of candidate nodes and results in $O(M\cdot d\cdot log(\gamma\cdot M))$ time.
Treating $M$ and $\gamma$ as constants, we see that at each visited node the computation complexity is $O(\log\gamma)$ and for greedy search at each level, the complexity is $O(\gamma\cdot \log(\gamma))$. Multiplying by $n\cdot \log(n)$ yields ACORN's final expected construction complexity, $O(n\cdot \gamma\cdot \log(n)\cdot \log(\gamma)$.

\subsection{Search Analysis}\label{subsec:discussion_search}
Turning our attention to ACORN-$\gamma$'s search algorithm, 
we will first point out several properties of HNSW that ACORN's predicate subgraphs aim to emulate. In Figure \ref{fig:sift_and_paper} we empirically show that ACORN's search performance approximates that of the HNSW oracle partition index.
We will then describe ACORN's expected search complexity. We define $l: X \to \mathbb{N}$ to be the mapping of nodes to there maximum level index in ACORN-$\gamma$.

\subsubsection{Index and Search Properties}
Intuitively, for a given query, ACORN's predicate subgraph will emulate the HNSW oracle partition index when the predicate subgraph forms a hierarchical structure, each node in the subgraph has degree close to $M$, the subgraph has a fixed entrypoint at its maximum level index that we can efficiently find during search, and the subgraph is connected. We will examine each of these properties separately and consider when they hold. We also note one main difference between ACORN's predicate subgraphs and HNSW that arises due to ACORN's predicate-agnostic pruning: each level of ACORN approximates a KNN graph, while each level of HNSW approximates a RNG graph. While this difference does not affect ACORN's expected search complexity in Section \ref{subsubsec:search_complexity}, Malkov et al. \cite{malkov_efficient_2018} demonstrated that the RNG-based pruning empirically improves performance.

\textbf{\emph{Hierarchy.}} First, we observe that the arbitrary predicate subgraph $G(X_p)$ forms a controllable hierarchy similar to the HNSW \emph{oracle partition index} built over $X_p$ with parameter $M$. This is by design. ACORN-$\gamma$'s construction fixes $M$, and consequently $m_L$, the level normalization constant. As a result, nodes of $X_p$ in the ACORN-$\gamma$ index are sampled at rates equal to the level probabilities of the HNSW partition. Ensuring this level sampling holds allows us to bound the expected greedy search path length at each level by a constant, $S$, as Malkov et al. \cite{malkov_efficient_2018} previously show. 

\textbf{\emph{Bounded Degree.}} Next, we will describe degree bounds, an important factor that impacts greedy search efficiency and convergence. While HNSW upper bounds the degree of each node by M \emph{during construction}, ACORN-$\gamma$ enforces this upper bound \emph{during search}. 
This ensures ACORN's search performs a constant number of distance computations per visited node. We now focus our attention on lower bounding the degree of nodes visited during ACORN-$\gamma$'s search over the predicate subgraph. 

If a node in the predicate subgraph has degree much lower than $M$, this could adversely impact the search convergence and thus recall. 
For a dataset and query predicate that exhibit no predicate clustering, for any node $v$ in $G(X_p)$, $$\mathbb{E} \bigr[|N_p^l(v)| \bigl] = |N^l(v)| \cdot s =  \gamma \cdot M \cdot s  > M, \forall s > s_{min}$$
This also holds as a lower bounds for datasets with predicate clustering, in which case $Pr(x \in N^l_p(v)) > s, \forall x \in N^l(v)$ where $v$ is a node in the predicate cluster. Thus we will continue our lower bound analysis of node degrees under the worst case assumption of no predicate clustering. Using the binomial concentration inequality with parameter $s$, and union-bounding over the expected search path length, we show that for the search path $\mathcal{P} = v_1 - ... - v_y$ over an arbitrary predicate subgraph:
\rev{$$ Pr\Bigl[\bigcup_{v\in \mathcal{P}} \bigl(|N_p(v)| \leq (1-\delta) M \bigr)\Bigr] \leq O\bigl(\log n \cdot \exp(-\delta^2 \gamma M  s /2)\bigr)$$}

We also analyze the probability that the subgraph traversal gets disconnected, which we bound by:
\rev{$$Pr\Bigl[\bigcup_{v\in \mathcal{P}} \bigl(|N_p(v)| \leq 0 \bigr) \Bigr] \leq O\bigl(\log n \cdot (1-s)^{M \cdot \gamma}\bigr)$$}
\rev{We see that both bounds decay exponentially in $\gamma$.}

\textbf{\emph{Fixed Entry-point.}} 
Similar to HNSW, ACORN's search begins from a fixed entry-point, chosen during construction. 
This pre-defined entry-point provides a simple and effective strategy that is also predicate-independent and robust to variations in \emph{query correlation}, as we empirically show in Figure \ref{fig2:laion1m_correlation}.

Intuitively, we expect the search to successfully navigate from ACORN's fixed entry-point, $e$, to the predicate-subgraph entry-point, $e_p$, when we find a node that passes the predicate on an upper level of the index that is fully connected. In this case, there will exist a one-hop path from $e$ to $e_p$. We consider $e_p$ to be an arbitrary node that passes a given predicate $p$ and is on the maximum level of the predicate subgraph. The index's neighbor expansion parameter, $\gamma$, causes the index's upper levels to be denser and, specifically those with less than $M\cdot \gamma$ nodes, to be fully connected. When these fully connected levels contain at least one node that passes the predicate, the search is guaranteed to route from $e$ to $e_p$. Since ACORN samples all nodes with equal probability at each level, the probability that nodes passing a given predicate, $p$, exist on some level is proportional to the predicate's selectivity, which takes a lower bound of $s_{min} = 1/\gamma$.

\textbf{\emph{Connectivity.}} 
\rev{We note that neither HNSW nor ACORN provides  theoretical guarantees on connectivity over its level graphs for arbitrary datasets. Thus we instead rely primarily on empirical results for our analysis. However, for some cases, we can expect ACORN's predicate subgraph to be connected when the HNSW oracle partition is connected. Two such cases are when $X_p$ exhibits no predicate clustering, or $X_p$ is clustered around a single region. In either case, each node has an expected degree of at least $M$ and each level approximates a KNN graph, which is connected when $K >> \log n$. 
We empirically show in Figure \ref{fig:num_scc} that ACORN's predicate subgraphs exhibit connectivity for real datasets and hybrid search queries. 
To analyze potential connectivity problems, we recommend bench-marking ACORN's hybrid search performance against HNSW's ANN search performance using equivalent $M$ and $\textit{efc}$ parameters. If a significant gap in accuracy exists, we recommend incrementally increasing $\gamma$ from its its initial value of $1/s_{min}$.}

\subsubsection{Search Complexity}\label{subsubsec:search_complexity}
ACORN-$\gamma$'s expected search complexity scales: $$O((d + \gamma) \cdot \log(s\cdot n) + \log(1/s))$$ 
This approximates the HNSW oracle partition's expected search complexity, $O(d \cdot \log(s\cdot n))$. Intuitively, ACORN-$\gamma$'s search path performs some filtering at the upper levels before likely entering and traversing the predicate sub-graph, during which ACORN incurs a small overhead compared to HNSW search in order to perform the predicate filtering step over each neighbor list.

We derive ACORN-$\gamma$'s search complexity by considering two stages of its search traversal. In the first stage, search begins from a pre-defined entry-point $e$, which need not pass the query predicate. In this stage, the search performs filtering only, dropping down each level on which the filtered neighbor list, $N_p(e)$, is found to be empty. Once the traversal reaches the first node, $e_p$ that passes the predicate, it enters the second stage, beginning its traversal over the predicate subgraph $G(X_p)$. 

In stage 1 the greedy search path on each layer has length 1, and occurs over $O(\log n - \log(s\cdot n))$ expected levels, yielding the complexity $O(\log(1/s)$. We see this because the expected maximum level index of the full ACORN index graph scales $O(\log n)$ based on its level-assignment probability \cite{malkov_efficient_2018}. Meanwhile, the predicate subgraph $G(X_p)$ of size $s\cdot n$ has an expected maximum level index of $O(\log(s\cdot n))$, once again according to its level sampling procedure.

The second stage of the search traverses the predicate subgraph in expected $O((d + \gamma)\cdot \log(s\cdot n))$ complexity. 
As we previously describe, the expected maximum level index of the predicate subgraph scales $O(\log(s\cdot n))$. 
At each level, the expected greedy path length can be bounded by a constant $S$ due to the index level sampling procedure employed during construction. For each node visited along the greedy path, we perform distance computations in $O(d)$ time on at most $M$ neighbors, and perform a constant-time predicate evaluations over at most $M\cdot \gamma$ neighbors.

%% file: Sections/Evaluation.tex
\section{Evaluation}\label{sec:eval}

\input{tables/datasets}

We evaluate ACORN through a series of experiments on real and synthetic datasets. Overall, our results show the following:
\begin{itemize}
    \item ACORN-$\gamma$ achieves state-of-the-art hybrid search performance, outperforming existing methods by 2-1,000$\times$ higher QPS at 0.9 recall on both prior benchmark datasets with simple, low-cardinality predicate sets, and more complex datasets with high-cardinality predicate sets. \rev{Specifically, ACORN achieves 2-10$\times$ higher QPS on prior benchmarks, over 30$\times$ higher QPS on new benchmarks, and over 1,000x higher QPS at scale on a 25-million-vector dataset.} 
    \item ACORN-$\gamma$ and ACORN-1 are predicate-agnostic methods, providing robust search performance under variations in predicate operators, predicate selectivity, query correlation, and dataset size.
    \item ACORN-1 and ACORN-$\gamma$ exhibit trade-offs between search performance and construction overhead. While ACORN-$\gamma$ achieves up to 5$\times$ higher QPS than ACORN-1 at fixed recalls, ACORN-1 can be constructed with 9-53$\times$ lower time-to-index (TTI).
\end{itemize}

We now discuss our results in detail. We first describe the datasets (\ref{subsec:Datasets}) and baselines (\ref{subsec:benchmarked-methods}) we use. 
Then, we present a systematic evaluation of ACORN's search peformance (\ref{subsec:search_perf}). 
Finally, we assess ACORN's construction efficiency (\ref{subsec:construction_oh}).
We run all experiments on an AWS m5d.24xlarge instance with 370 GB of RAM,  96 vCPUs, \rev{and 196 threads}.

\subsection{Datasets}\label{subsec:Datasets}
\footnotetext{On the TripClick dataset, we create two distinct query workloads, described in Section \ref{subsubsec:hcps_datasets}. The average selectivity for either workload is .17 (keywords), and .26 (dates).}
\footnotetext{On the LAION dataset, we create four distinct query workloads, described in Section \ref{subsubsec:hcps_datasets}. These workloads have average selectivities of .10 (no-cor), .13 (pos-cor), .069 (neg-cor), .056 (regex).}

We conduct our experiments on two datasets with low-cardinality predicate sets (LCPS) and two datasets with high-cardinality predicate sets (HCPS). 
The LCPS datasets allow us to benchmark prior works that only support a constrained set of query predicates. 
The HCPS datasets consist of more complex and realistic query workloads, allowing us to more rigorously evaluate ACORN's search performance.
Table \ref{tab:dataset} provides a concise summary of all datasets.

\subsubsection{Datasets with Low Cardinality Predicate Sets}\label{subsubsec:lcps_dataset} 
We use SIFT1M \cite{jegou_product_2011} and Paper \cite{wang_navigable_2022}, the two largest publically-available datasets used to evaluate recent specialized indices \cite{wang_navigable_2022, gollapudi_filtered-diskann_2023}.
For both datasets, we follow related works \cite{gollapudi_filtered-diskann_2023, wang_navigable_2022, wangmilvus2021} to generate structured attributes and query predicates: for each base vector, we assign a random integer in the range $1-12$ to represent structured attributes; and for each query vector, the associated query predicate performs an exact match with a randomly chosen integer in the attribute value domain. The resulting query predicate set has a cardinality of 12.

\underline{SIFT1M:} The SIFT1M dataset was introduced by Jegou et al. in 2011 for ANN search. It consists of a collection of 1M base vectors, and 10K query vectors. All of the vectors are 128-dimensional local SIFT descriptors \cite{lowe_distinctive_2004} from INRIA Holidays images \cite{jegou_hamming_2008}.

\underline{Paper:} Introduced by Wang et al. in 2022, the Paper dataset consists of about 2M base vectors and 10K query vectors. The dataset is generated by extracting and embedding the textual content from an in-house corpus of academic papers.

\subsubsection{Datasets with High Cardinality Predicate Sets}\label{subsubsec:hcps_datasets} We use the TripClick and LAION datasets in our experiments with HCPS datasets. 

\underline{TripClick:} The TripClick dataset, introduced by Rekabsaz et al. in 2021 for text retrieval, contains a \emph{real} hybrid search query workload and base dataset from the click logs of a health web search engine. \rev{Each query consists of natural language search terms along with optional filters on clinical areas (e.g. "cardiology", "infectious disease", "surgery") and publication years. Each entity in the base dataset consists of a text passage, with a list of associated clinical areas and a publication date. The dataset contains 28 unique clinical areas and publication dates ranging from 1900 to 2020, resulting in over $2^{28}$ possible query predicates total. We construct two query workloads, one consisting of queries that used date filters (dates) and another consisting of queries that used clincal area filters (areas). We generate 768-dimensional vectors from the query texts and passage texts using DPR \cite{karpukhin_dense_2020}, a widely-used, pre-trained encoder for open-domain Q\&A. The resulting dataset has about $1M$ base vectors, and we use a random sample of 1K queries for each query workload.} 

\underline{LAION:} The LAION dataset \cite{schuhmann_laion-400m_2021} consists of 400M image embeddings and captions describing each image. The vector embeddings are generated from web-scraped images using CLIP \cite{radford_learning_2021}, a multi-modal language-vision model. In our evaluation, we construct two base datasets using 1M and 25M LAION subsets, both consisting of image vectors and text captions as a structured attribute. We also generate an additional structured attribute consisting of a keyword list. We assign each image embedding its keyword list by taking the 3 words with highest text-to-image CLIP scores from a candidate list of 30 common adjectives and nouns (e.g., "animal", "scary").
    
\begin{figure*}[!t]
  \centering
  \includegraphics[trim={0.0in 0 0.00in 0.0in 0.0in}, clip, width=0.7\linewidth]{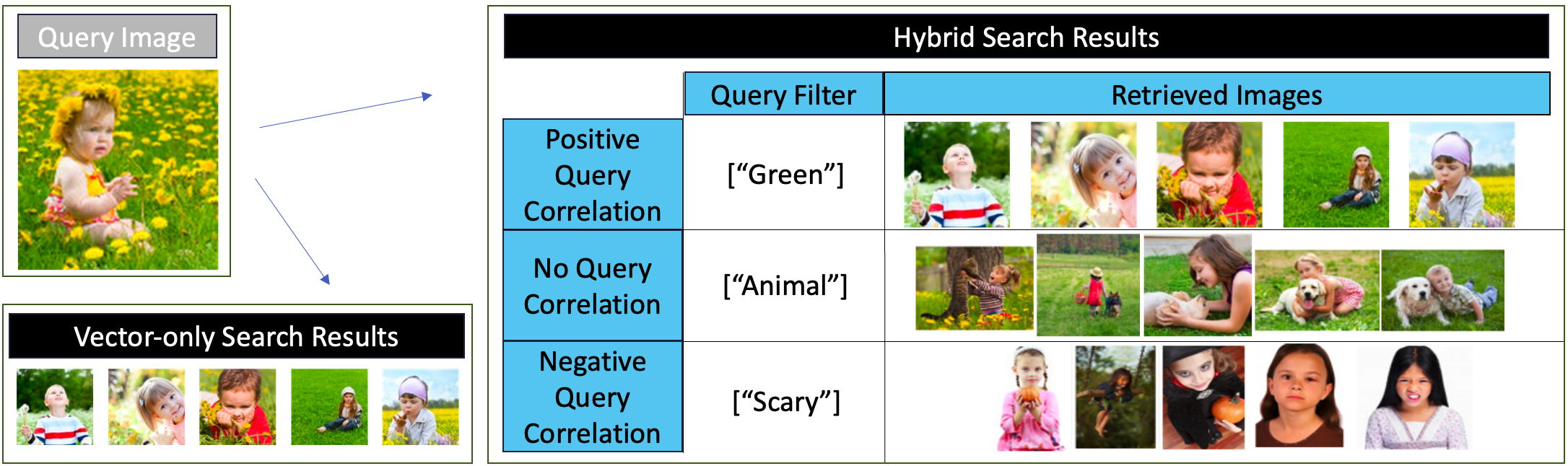}
  \caption{\small The figure contrasts retrieval results using vector-only similarity search (bottom left) versus hybrid search (right) on the LAION dataset. Both use the same query image (top left), and the hybrid search queries also include a structured query filter consisting of a keyword list, here containing a single keyword. The table on the right shows examples from three hybrid search query workloads: positive query correlation (top), no query correlation (middle), and negative query correlation (bottom). }
  \vspace{-.39cm}
  \label{fig:Sample laion query and results}
\end{figure*}

To evaluate a series of micro-benchmarks, we generate four query workloads. For each query workload, we sample 1K vectors from the dataset as query vectors. We construct the regex query workload with predicates that perform regex-matching over the image captions. For each query predicate, we randomly choose strings of 2-10 regex tokens (e.g., "\verb|^[0-9]|"). In addition, we construct three query workloads with predicates, similar to TripClick, that take a keyword list and filter out entities that do not have at least one matching keyword. Using this setup, we are able to easily control for correlation in the workload, and we generate a no correlation (no-cor), positive correlation (pos-cor), and negative correlation (neg-cor) workload. Figure \ref{fig:Sample laion query and results} demonstrates some example queries and multi-modal retrieval results taken from each.

\subsection{Benchmarked Methods}\label{subsec:benchmarked-methods}
We briefly overview the methods we benchmark along with tested parameters. We implement ACORN-$\gamma$, ACORN-1, pre-filtering, and HNSW post-filtering in C++ in the FAISS codebase \cite{noauthorfaiss2023}.

\underline{HNSW Post-filtering:}
To implement HNSW post-filtering, for each hybrid query with predicate selectivity $s$, we over-search the HNSW index, gathering $K / s$ candidate results before applying the query filter. We note that this differes to some prior work \cite{gollapudi_filtered-diskann_2023}, where HNSW post-filtering is implemented by collecting only $K$ candidate results, leading to significantly worse baseline query performance than ours. 
For the SIFT1M, Paper and LAION datasets, we use the FAISS's default HNSW construction parameters: $M=32, \textit{efc}=40$. For the TripClick dataset, we find that the HNSW index for these parameters is unable to obtain high recalls for the standard ANN search task, thus we perform parameter tuning, as is standard. We perform a grid search for $M \in \{32, 64, 128\}$ and $\textit{efc} \in \{40, 80, 120, 160, 200\}$ and choose the pair the obtains the highest QPS at 0.9 Recall for ANN search. For TripClick, we choose $M=128, \textit{efc}=200$. We generate each recall-QPS curve by varying the search parameter $\textit{efs}$ from 10 to 800 in step sizes of 50.

\underline{Pre-filtering:} \rev{We implement pre-filtering by first generating a list of dataset entries that pass the query predicate and then performing brute force search using FAISS's optimized implementation for distance comparisons. 
We also efficiently implement all \texttt{contains} predicate evaluations using bitsets since the corresponding structured attributes have low cardinality.
}

\underline{Filtered-DiskANN:}
We evaluate both algorithms implemented in FilteredDiskANN \cite{noauthor_diskann_2023}, namely FilteredVamana and StitchedVamana. For both, we follow the recommended construction and search parameters according to the hyper-parameter tuning procedure described by Gollapudi et al. \cite{gollapudi_filtered-diskann_2023}. For FilteredVamana, we use construction parameters $L=90, R=96$, which generated the Pareto-Optimal recall-QPS curve from a parameter sweep over $R\in \{32, 64, 96\}$ and L between 50 and 100. For StitchedVamana, we use construction parameters $R_{small}=32, L_{small}=100, R_{stitched}=64$ and $\alpha=1.2$, which generated the Pareto-Optimal recall-QPS curve from a parameter sweep over $R_{small}, R_{stitched} \in \{32, 64, 96\}$ and $L_{small}$ between 50 and 100. To generate the recall-QPS curves we vary $L$ from 10 to 650 in increments of 20 for FilteredVamana, and $L_{small}$ from 10 to 330 in increments of 20 for StitchedVamana.

\underline{NHQ}:
 We evaluate the two algorithms, NHQ-NPG\_NSW and NHQ-NPG\_KGraph, proposed in \cite{wang_navigable_2022}. For both we use the recommended parameters in the released codebase \cite{ashenon3_nhq_2023}. These parameters were selected using a hyperparameter grid search in order to generate the Pareto-optimal recall-QPS curve for either algorithm on the SIFT1M and Paper datasets. We generate the recall-QPS curve by varying $L$ between 10 and 310 in steps of 20. In Figures \ref{fig1a:query_perf_sift} and \ref{fig1b:query_perf_Paper}, we show the query performance of KGraph, the more performant of the two algorithms.

\begin{figure*}
    \centering
    \vspace{-.5cm}
    \includegraphics[trim={0.0in 3.20in 0.0in 0.0in}, clip, width=0.6\linewidth]{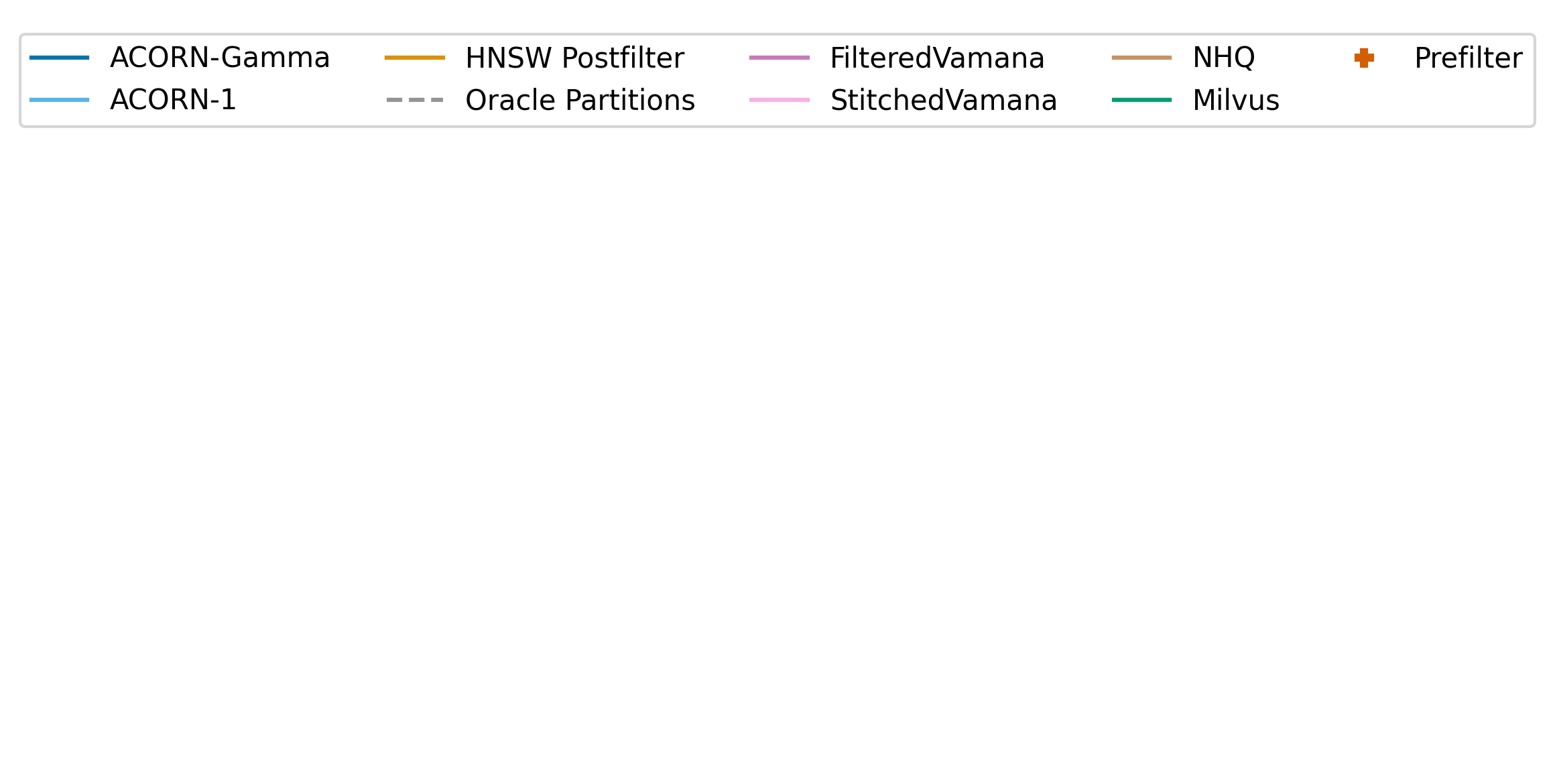}
    \label{fig:legend_large}
\end{figure*}

\begin{figure}[!th]
  \centering
  \vspace{-.5cm}
  \begin{subfigure}{0.4\linewidth}
    \centering
    \includegraphics[width=\linewidth]{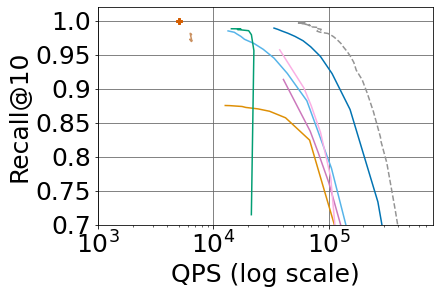}
    \caption{SIFT1M Dataset}
    \label{fig1a:query_perf_sift}
  \end{subfigure}
  \begin{subfigure}{0.4\linewidth}
    \centering
    \includegraphics[width=\linewidth]{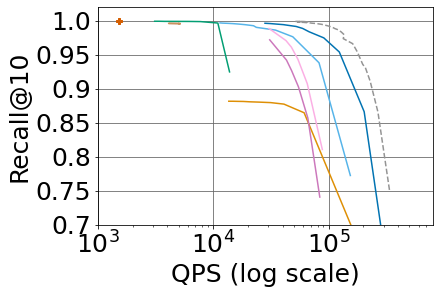}
    \caption{Paper Dataset}
    \label{fig1b:query_perf_Paper}
  \end{subfigure}
  \vspace{-.3cm}
  \caption{\small Recall@10 vs QPS on SIFT1M and Paper}
  \label{fig:sift_and_paper}
   \vspace{-.5cm}
\end{figure}

\underline{Milvus:}
We test four Milvus algorithms: IVF-Flat, IVF-SQ8, HNSW, and IVF-PQ \cite{noauthor_milvus_2023}. For each we test the same parameters as Gollapudi et al. \cite{gollapudi_filtered-diskann_2023}. Since we find that the four Milvus algorithms achieve similar search performance, for simplicity, Figures \ref{fig1a:query_perf_sift} and \ref{fig1b:query_perf_Paper} show only the method with Pareto-Optimal recall-QPS performance.

\underline{Oracle Partition Index}: We implement this method by constructing an HNSW index for each possible query predicate in the LCPS datasets. For a given hybrid query, we search the HNSW partition corresponding to the query's predicate. To construct each HNSW partition and generate the recall-QPS curve, we use the same parameters as the HNSW post-filtering method, described above.

\underline{ACORN-$\gamma$}:
We choose the construction parameters $M$ and $\textit{efc}$ to be the same as the HNSW post-filtering baseline, described above. We find that ACORN-$\gamma$'s search performance is relatively in-sensitive to the choice of the construction parameter $M_{\beta}$, as Figure \ref{fig:pruning_comparison}c shows. 
Thus, to maintain modest construction overhead, we choose $M_\beta$ to be a small multiple of $M$, \emph{i.e.,} $M_{\beta}=M$ or $M_{\beta}=2M$, picking the parameter for each dataset that obtains higher QPS at 0.9 Recall.
\rev{Specifically, we constrain the memory budget of the index to be no larger than the Vamana indices on the LCPS datasets and no larger than twice the size of the flat indices for HCPS datasets.
We use $M_\beta$ values of $32$ for LAION-1M and LAION-25M, $64$ for SIFT1M, Paper, and $128$ for TripClick. }
We choose the construction parameter $\gamma$ according to the expected minimum selectivity query predicates of each dataset \emph{i.e.,} $\gamma=12$ for SIFT1M and Paper, $\gamma=30$ for LAION, and $\gamma=80$ for TripClick. To generate the recall-QPS curve, we follow the same procedure described above for HNSW post-filtering.

\underline{ACORN-1:}
We construct ACORN-$1$ and generate the recall-QPS curve following the same procedure we use for ACORN-$\gamma$, except that we fix $\gamma=1$ and $M_{\beta} = M$.

\subsection{Search Performance Results}\label{subsec:search_perf}
We will begin our evaluation with benchmarks on the LCPS datasets, on which we are able to run all baseline methods as well as the oracle partition method. We will then present an evaluation on the HCPS datasets. On these datasets, the FilteredDiskANN and NHQ algorithms fail because they assume are unable to handle the high cardinality query predicate sets and non-equality predicate operators. As of this writing, we also find that Milvus cannot support \verb|regex-match| predicates and \verb|contains| predicates over variable length lists. As a result, we instead focus on comparing ACORN to the pre- and post-filtering baselines for the HCPS datasets. We report QPS averaged over 50 trials.

\subsubsection{Benchmarks on LCPS Datasets}\label{subsubsec:search_perf_sift_and_paper} Figure \ref{fig:sift_and_paper} shows that ACORN-$\gamma$ achieves state-of-the-art hybrid search performance and best approximates the theoretically ideal \emph{oracle partition strategy} on the SIFT1M and Paper datasets.
Notably, even compared to NHQ and FilteredDiskANN, which specialize for LCPS datasets, ACORN-$\gamma$ consistently achieves 2-10$\times$ higher QPS at fixed recall values, while maintaining generality. 
Additionally, we see ACORN-1 approximates ACORN-$\gamma$'s search performance, attaining about 1.5-5$\times$ lower QPS than ACORN-$\gamma$ across a range of recall values.

\input{tables/distance_comps}
To further investigate the relative search efficiency of ACORN-$\gamma$ and ACORN-$1$, we turn our attention to Table \ref{tab:dist_comps}, which shows the number of distance computations required of either method to obtain Recall@10 equal to 0.8.  We see that the oracle partition method is the most efficient, requiring the fewest number of distance computations on both datasets. ACORN-$\gamma$ is the next most efficient according to number of distance computations. 
While ACORN-$\gamma$ approximates the oracle partition method, it's predicate-agnostic design precludes the same RNG-based pruning used to construct the oracle partitions. Rather than approximating RNG-graphs, ACORN-$\gamma$'s levels approximate KNN-graphs, which are less efficient to search over explaining the performance gap.
The table additionally shows that ACORN-1 is less efficient than ACORN-$\gamma$, which is explained by the candidate edge generation used in ACORN-1. While the ACORN-$\gamma$ index stores up to $M\times \gamma$ edges per node during construction, ACORN-1 stores only up to $M$ edges per node during construction, and \emph{approximates} an edge list of size $M*\gamma$ for each node during search using its neighbor expansion strategy.
This approximation results in slight degradation to neighbor list quality and thus search performance.
Finally, we see from the table, that HNSW post-filtering is the least efficient of the listed methods. 
This is because while  ACORN-1 and ACORN-$\gamma$ almost exclusively traverse over nodes that pass the query predicates, the post-filtering algorithm is less discriminating and wastes distance computations on nodes failing the query predicate.

Returning to Figure \ref{fig:sift_and_paper}, we see that the relative search efficiency, measured by QPS versus recall, of the oracle partition method, ACORN-$\gamma$, and ACORN-1 is not only affected by distance computations, but is also affected by vector dimensionality. We see that both ACORN-1 and ACORN-$\gamma$ perform closer to the oracle partition method on the Paper dataset, while the performance gap grows slightly on SIFT1M. This is due to the cost of performing the filtering step over neighbor lists during search, which, relative to the cost of distance computations, is higher on SIFT1M than Paper since SIFT1M uses slightly lower-dimensional vectors.

\subsubsection{Benchmarks on HCPS Datasets}
\begin{figure}[!t]
  \centering
  \vspace{-.5cm}
  \label{fig:legend_small}
  \begin{subfigure}{0.32\linewidth}
    \centering
    \includegraphics[width=\linewidth]{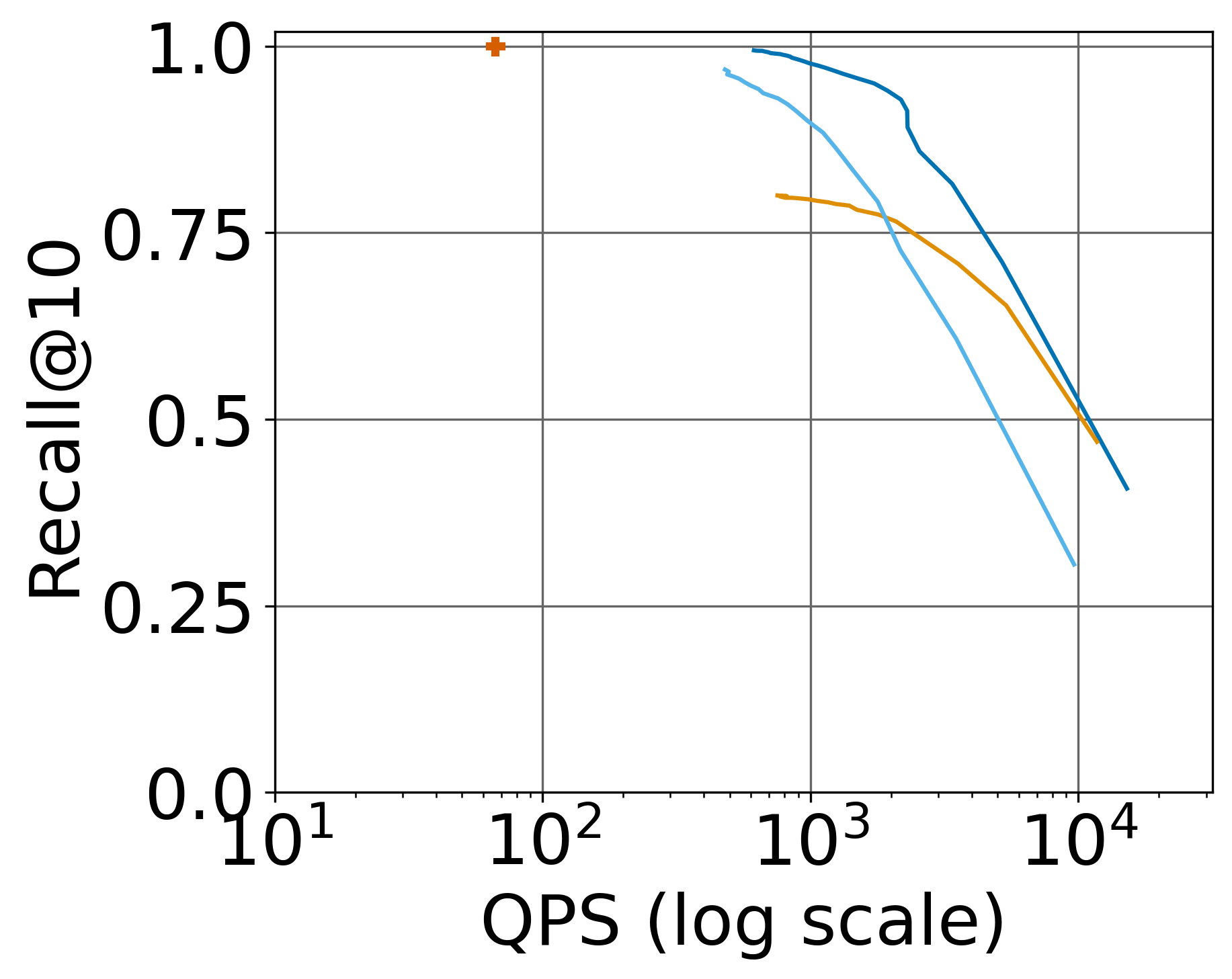}
    \caption{TripClick (areas)}
    \label{fig1a:query_perf_sift}
  \end{subfigure}
  \begin{subfigure}{0.32\linewidth}
    \centering
    \includegraphics[width=\linewidth]{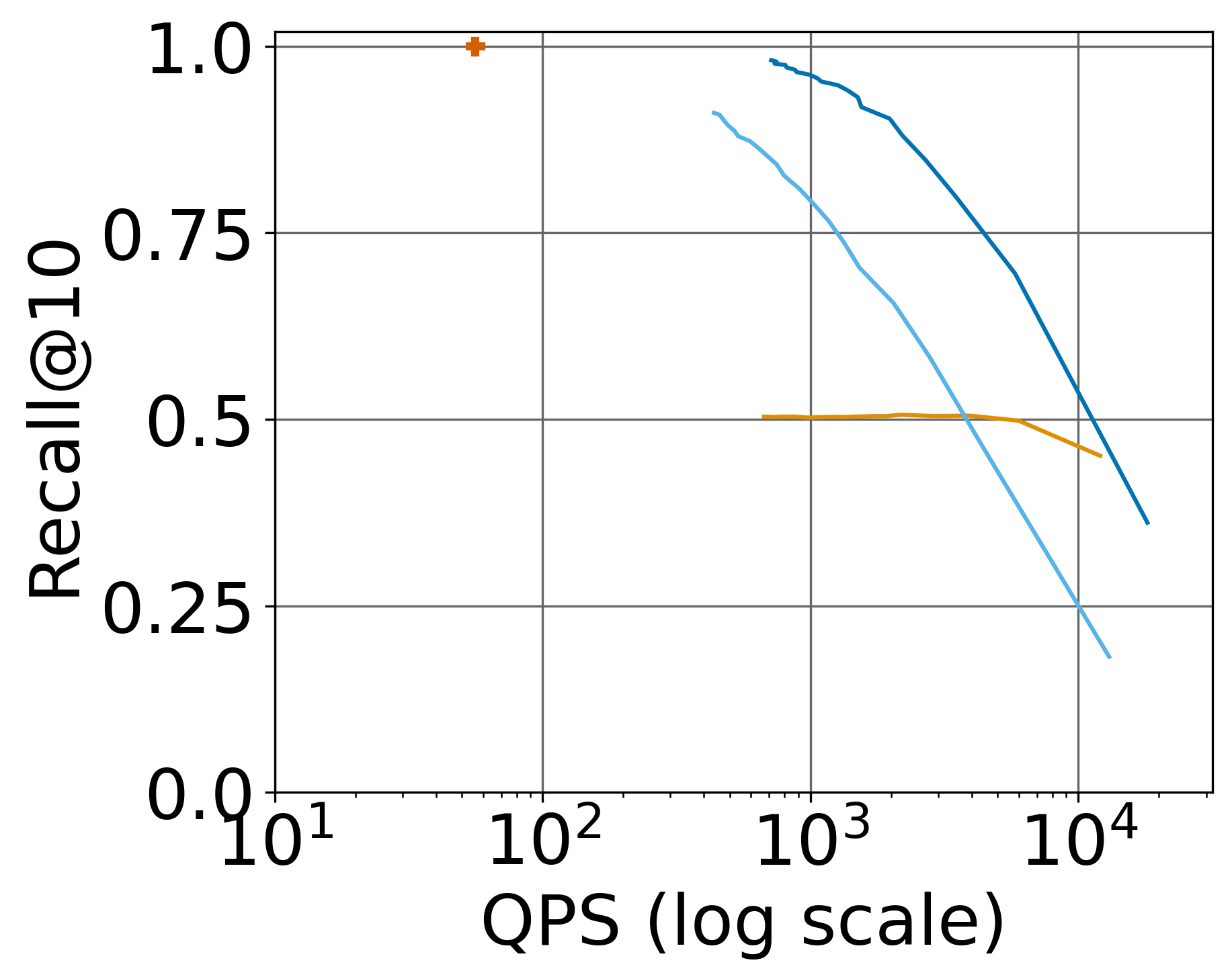}
    \caption{\rev{TripClick (dates)}}
    \label{fig1a:query_perf_sift}
  \end{subfigure}
  \begin{subfigure}{0.32\linewidth}
    \centering
    \includegraphics[width=\linewidth]{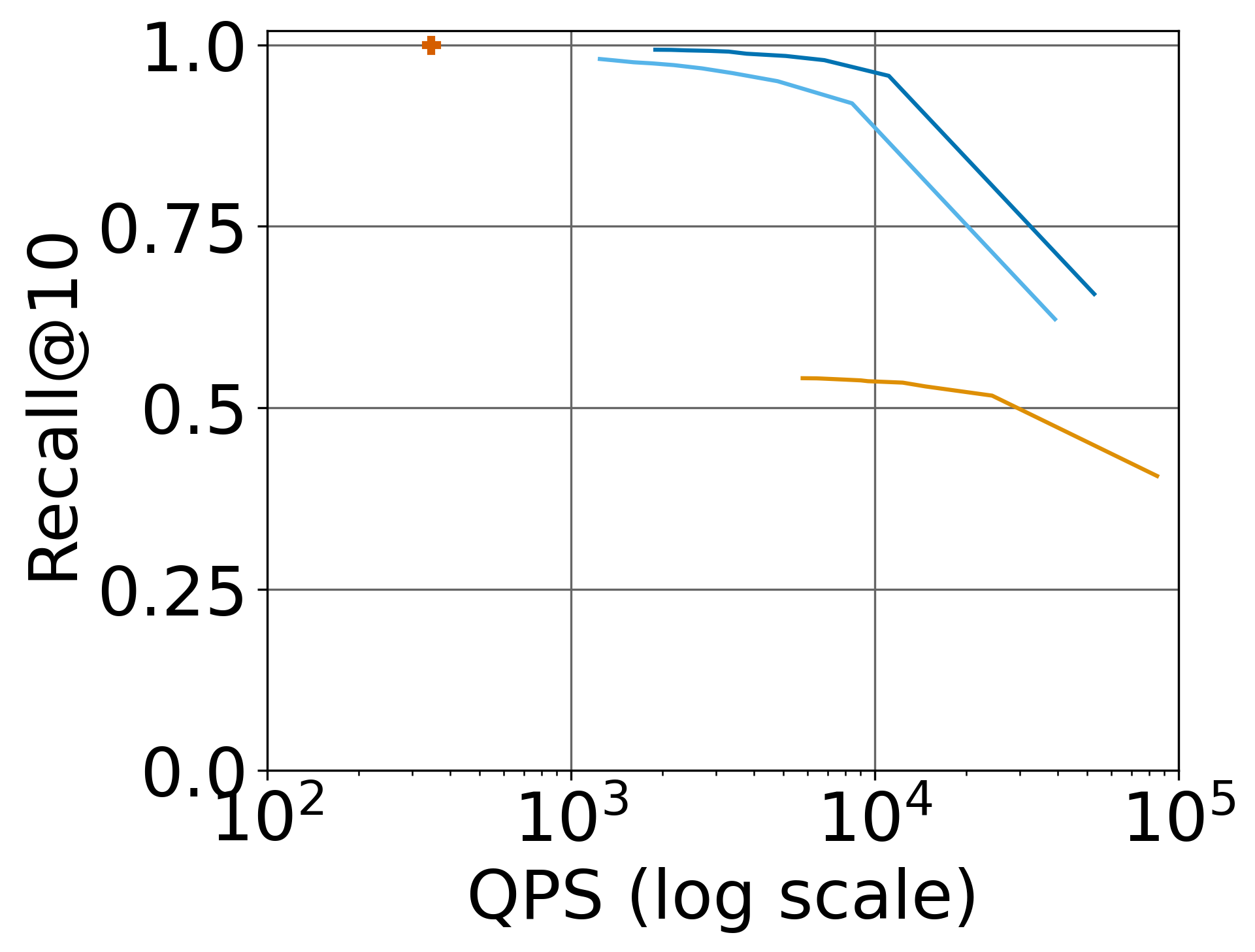}
    \caption{LAION1M (regex)}
    \label{fig1b:query_perf_tripclick}
  \end{subfigure}
  \vspace{-.3cm}
  \caption{\small Recall@10 vs QPS on TripClick and LAION-1M}
  \label{fig:tripclick_and_laion}
  \vspace{-.5cm}
\end{figure}

Figure \ref{fig:tripclick_and_laion} shows that ACORN outperforms the baselines by $30 - 50\times$ higher QPS at 0.9 recall on TripClick and LAION-1M, and as before, ACORN-1 approximates ACORN-$\gamma$'s search performance. On both datasets, pre-filtering is prohibitively expensive, obtaining perfect recall at the price of efficiency. Meanwhile, post-filtering fails to obtain high recall, likely due to the presence of varied query correlation and predicate selectivity, which we further explore further next.

\begin{figure*}[!ht]
  \centering
  \begin{subfigure}{0.18\linewidth}
    \centering
    \includegraphics[width=\linewidth]{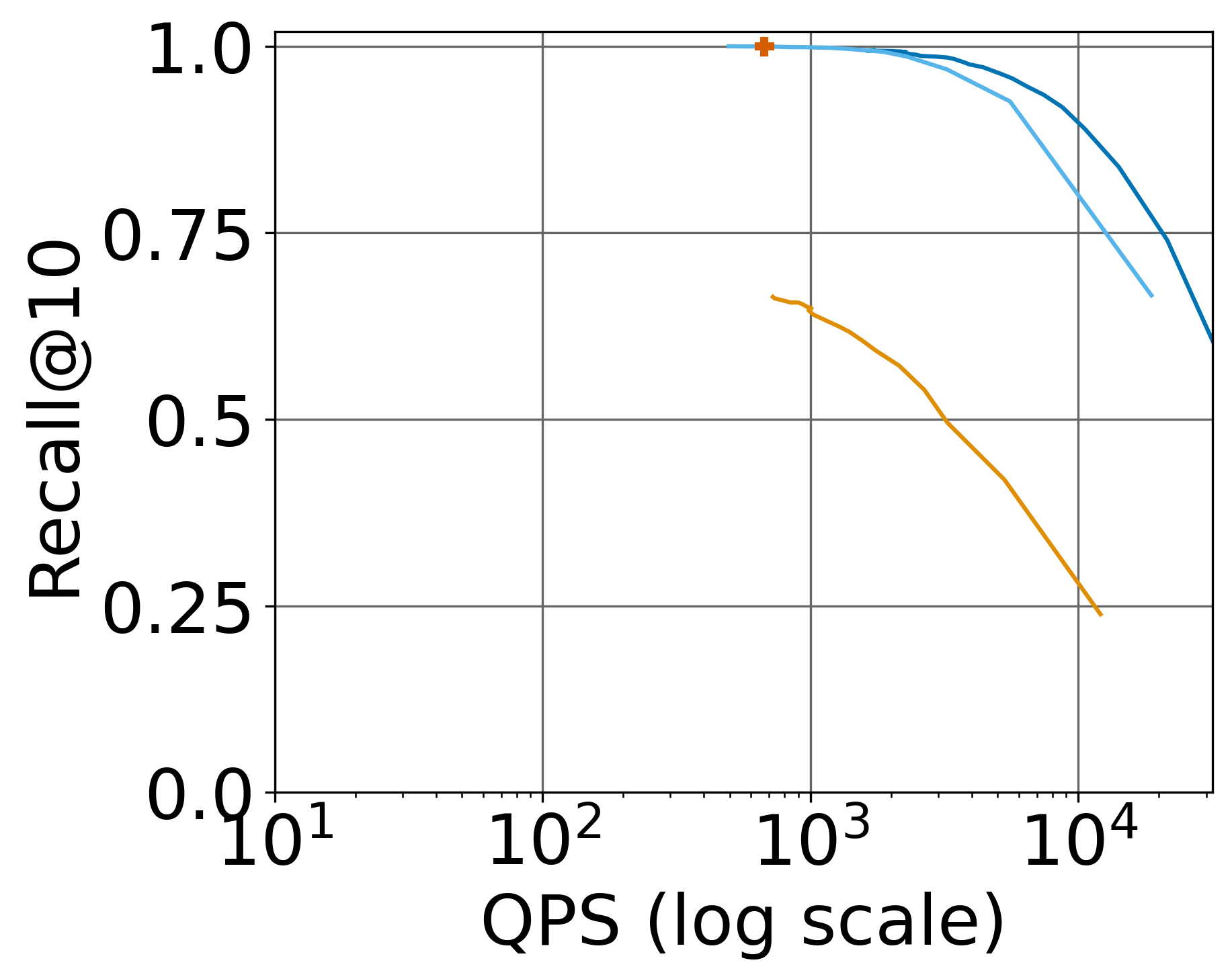}
    \caption{1p Sel (s=0.0127)}
    \label{tripclick_1p}
  \end{subfigure}
  \hfill
  \begin{subfigure}{0.18\linewidth}
    \centering
     \includegraphics[width=\linewidth]{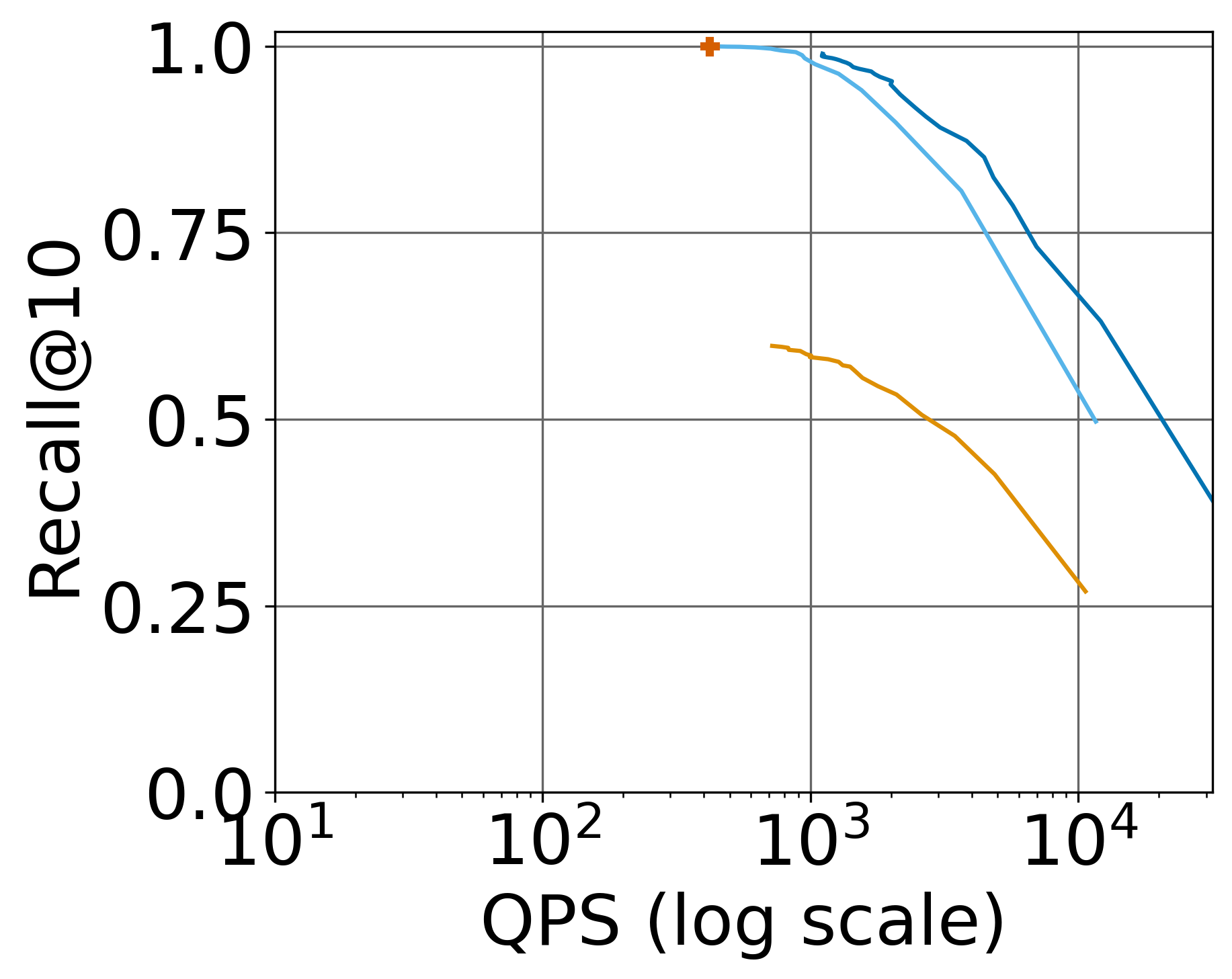}
    \caption{25p Sel (s=0.0485)}
    \label{tripclick_25p}
  \end{subfigure}
  \hfill
  \begin{subfigure}{0.18\linewidth}
    \centering
     \includegraphics[width=\linewidth]{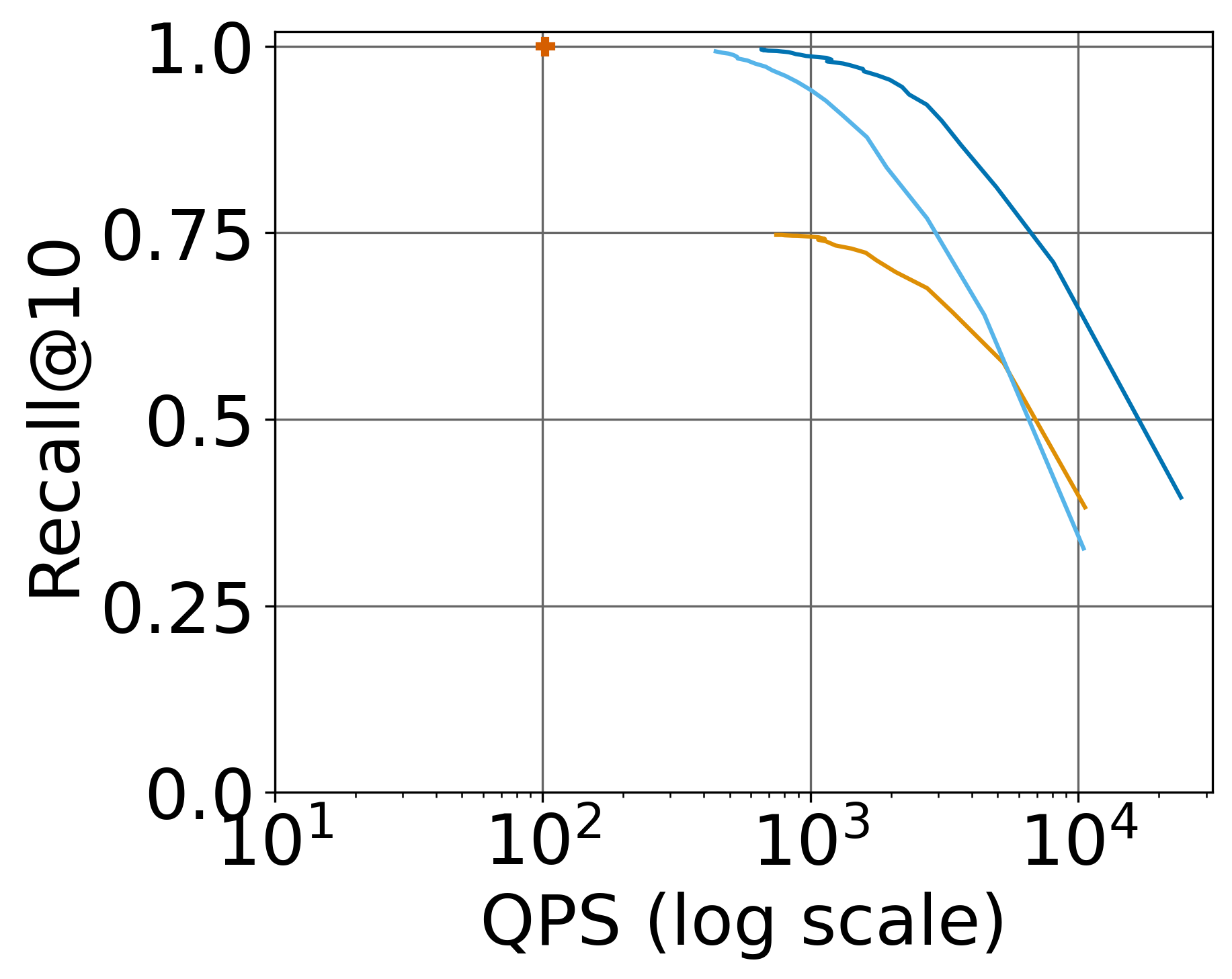}
    \caption{50p Sel (s=0.1215)}
    \label{tripclick_50p}
  \end{subfigure}
  \hfill
  \begin{subfigure}{0.18\linewidth}
    \centering
     \includegraphics[width=\linewidth]{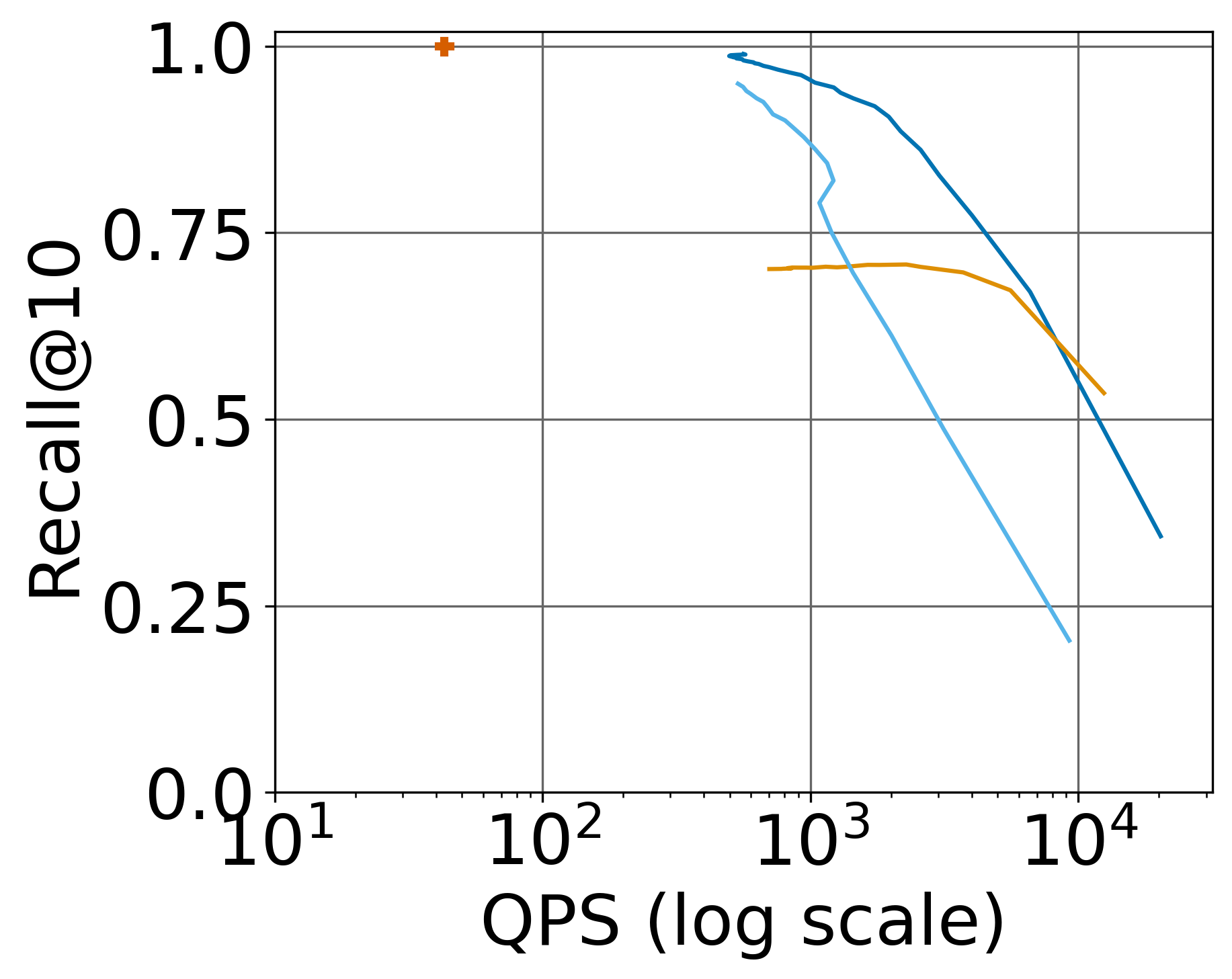}
    \caption{75p Sel (s=0.2529)}
    \label{tripclick_75p}
  \end{subfigure}
  \begin{subfigure}{0.18\linewidth}
    \centering
     \includegraphics[width=\linewidth]{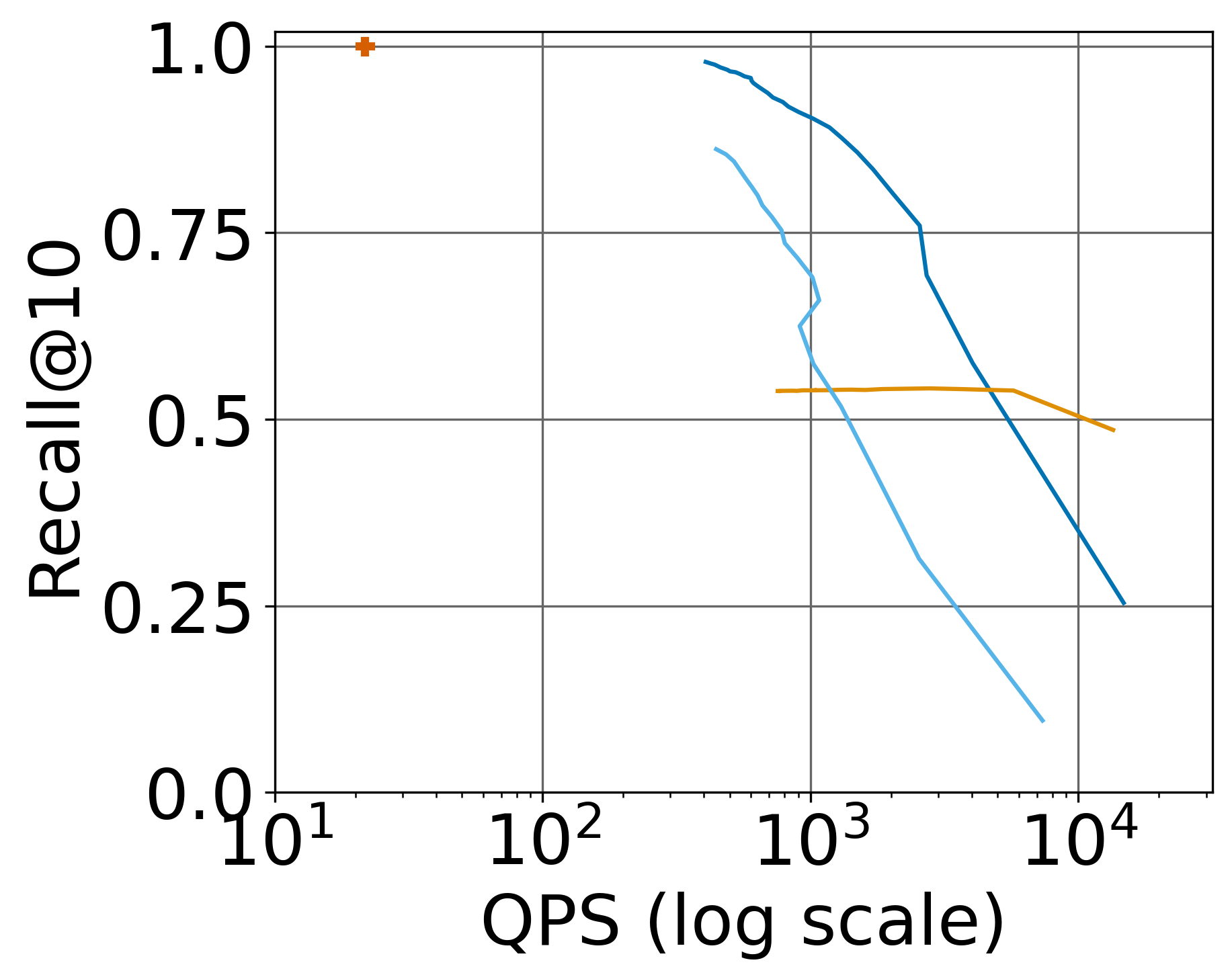}
    \caption{99p Sel (s=0.6164)}
    \label{tripclick_99p}
  \end{subfigure}  
  \vspace{-.2cm}
  \caption{\small Recall@10 vs QPS for Varied Selectivity Query Filters on TripClick}
  \label{fig:tripclick}
\end{figure*}

\underline{Varied Predicate Selectivity:}\label{subsubsec:search_sel} 
We use the Tripclick dataset to evaluate ACORN's search performance across a range of realistic predicate selectivities. 
Figure \ref{fig:tripclick} demonstrates that for each predicate selectivity percentile, ACORN-$\gamma$ achieves 5-50x higher QPS at 0.9 recall compared to the next best-performing baseline. Once again ACORN-1 trails behind ACORN-$\gamma$. We see that for low selectivity predicates, the pre-filtering method is most competitive, while the post-filtering baselines suffers from over 10$\times$ lower QPS than ACORN at fixed recall. However, for high selectivity predicates, pre-filtering becomes less competitive while the post-filtering baseline obtains higher throughput, although its recall remains low.

\begin{figure}[!tb]
\vspace{-.37cm}
  \centering
  \begin{subfigure}{0.30\linewidth}
    \centering
    \includegraphics[width=\linewidth]{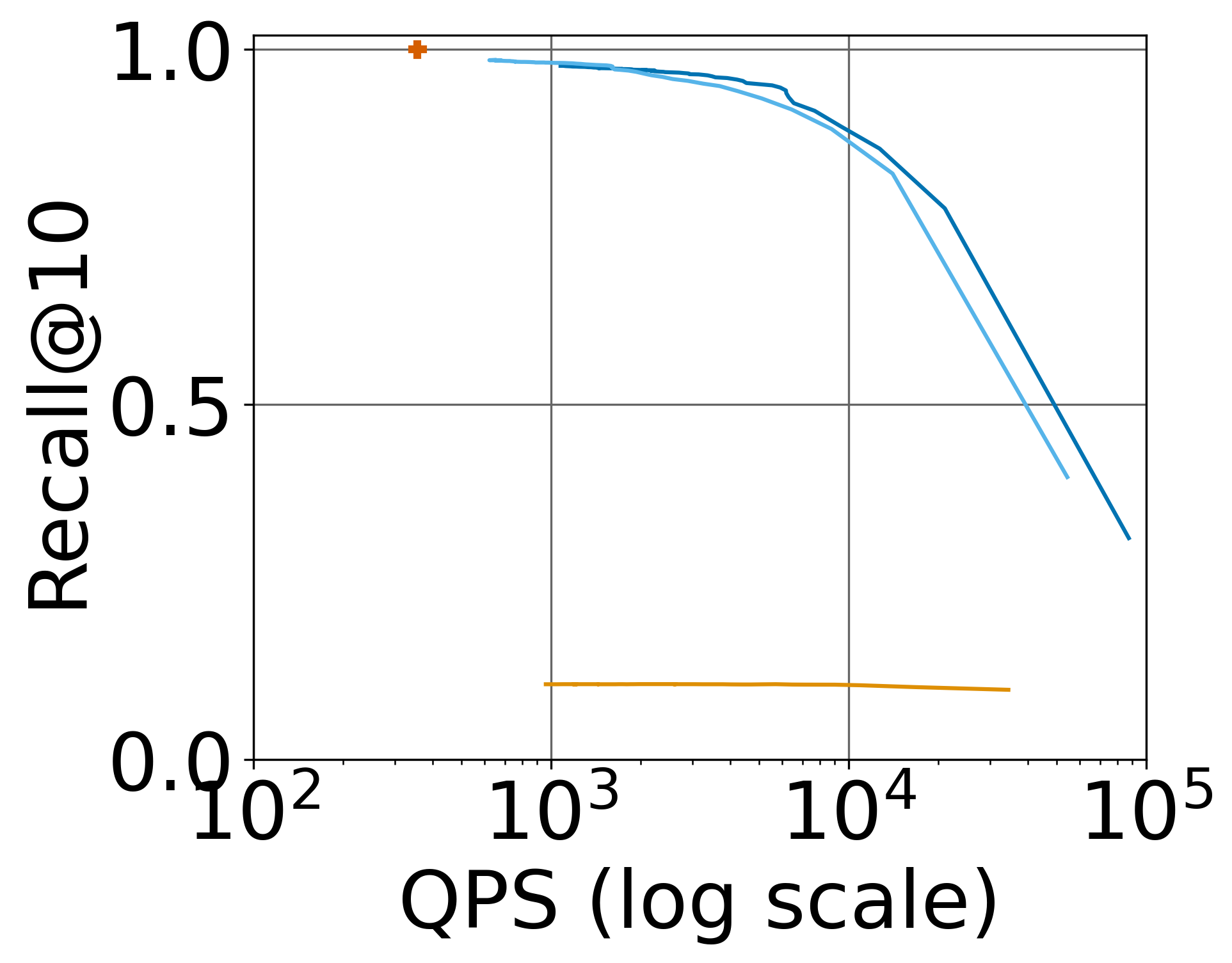}
    \caption{Neg. Correlation}
    \label{laion_negcorr}
  \end{subfigure}
  \hfill
  \begin{subfigure}{0.30\linewidth}
    \centering
    \includegraphics[width=\linewidth]{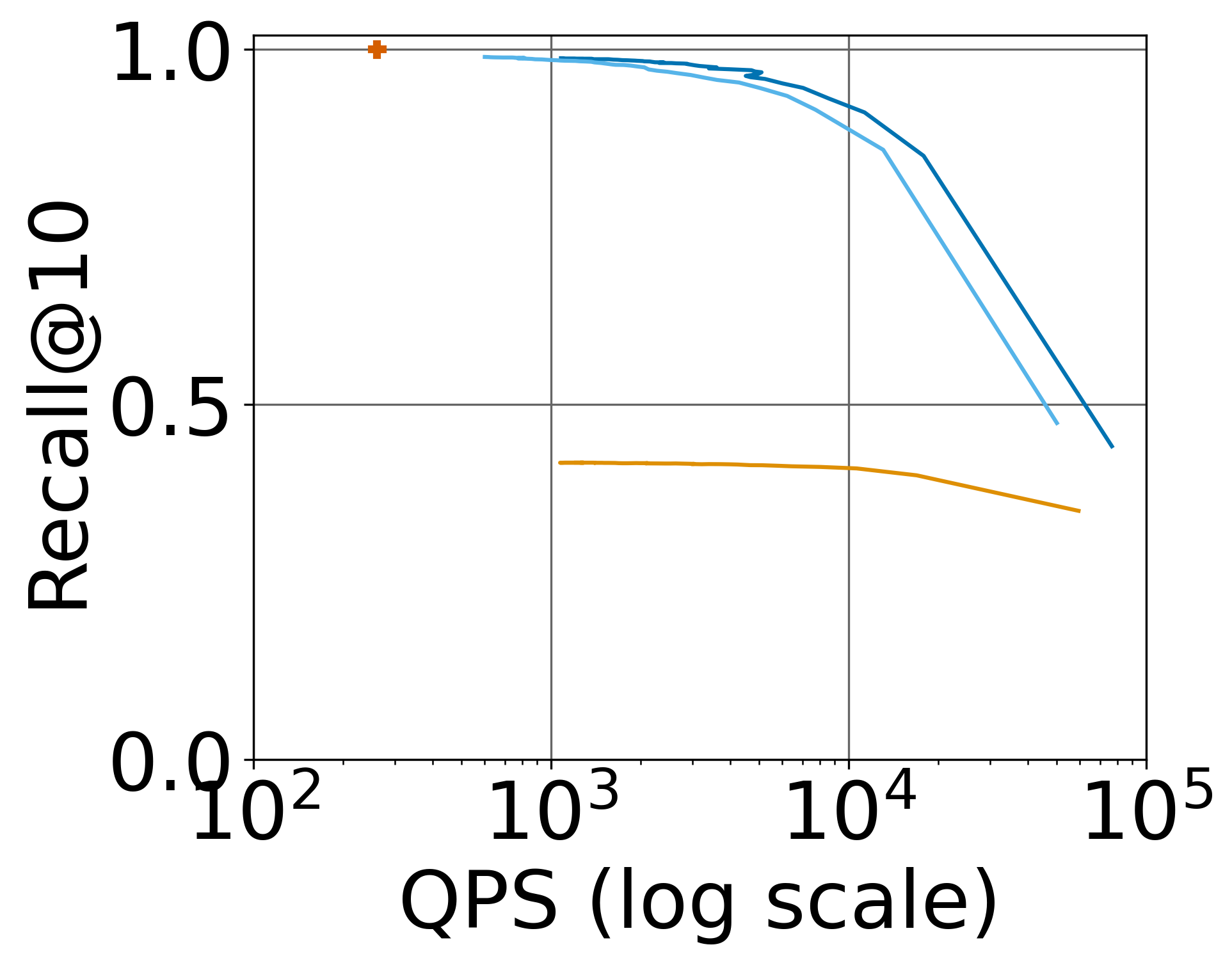}
    \caption{No Correlation}
    \label{laion_noscorr}
  \end{subfigure}
  \hfill
  \begin{subfigure}{0.30\linewidth}
    \centering
    \includegraphics[width=\linewidth]{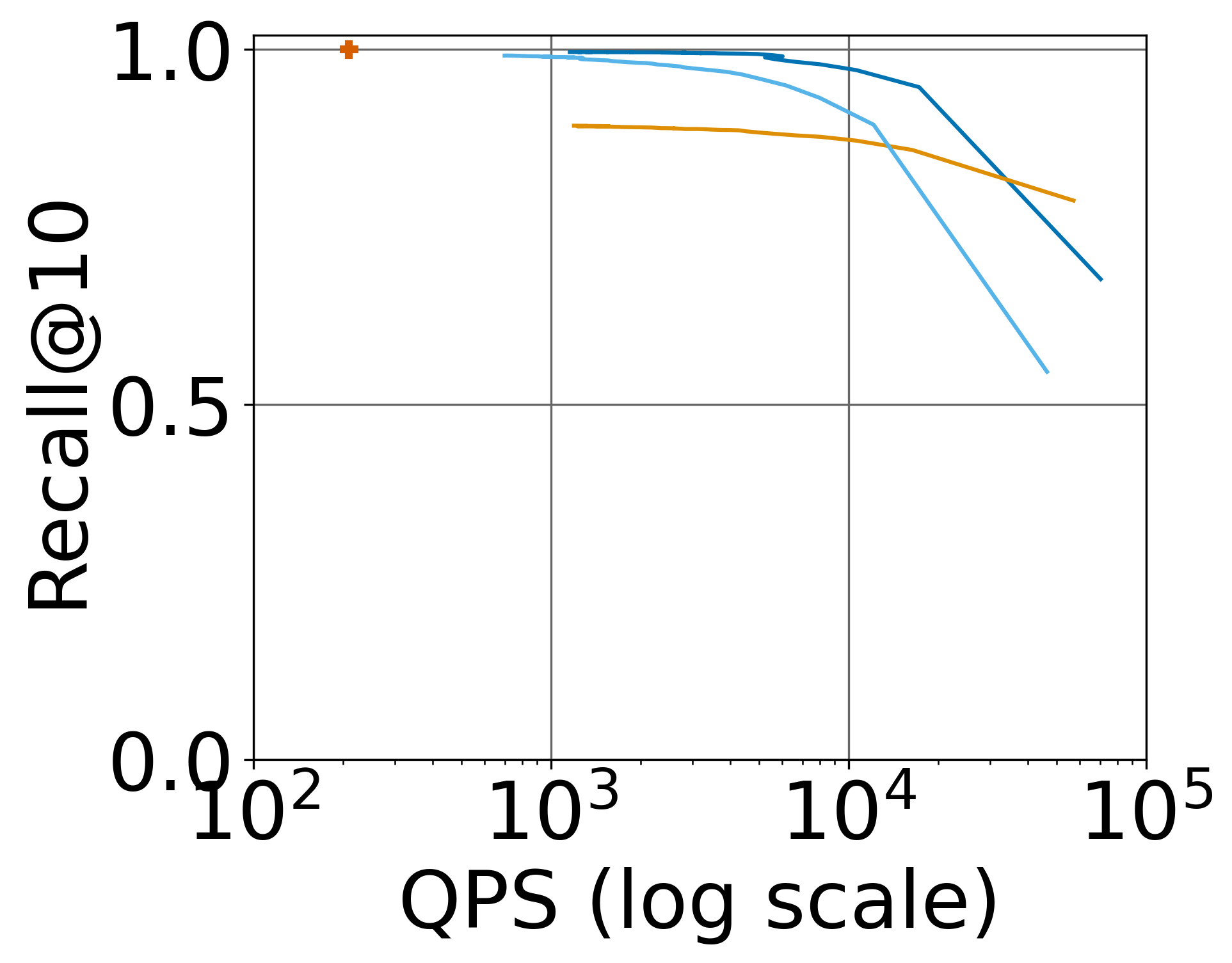}
    \caption{Pos. Correlation}
    \label{laion_poscorr}
  \end{subfigure}
  \vspace{-.2cm}
  \caption{\small Recall@10 vs QPS on LAION1M}
  \vspace{0cm}
  \label{fig2:laion1m_correlation}
\end{figure}

\underline{Varied Query Correlation:}\label{subsubsec:search_corr} 
Next we control for query correlation and evaluate ACORN on three different query workloads using the LAION-1M dataset. Figure \ref{fig2:laion1m_correlation} demonstrates that ACORN-$\gamma$ is robust to variations in query correlation and attains 28-100$\times$ higher QPS at 0.9 recall than the next best baseline in each case. In the negative correlation case, the performance gap between post-filtering and the ACORN methods is the largest since post-filtering cannot successfully route towards nodes that pass the predicate. In the positive correlation case, ACORN-$\gamma$ once again outperforms the baselines, but post-filtering become more competitive, although it is still unable to attain recall above 0.9. The pre-filtering method's QPS remains relatively unchanged, and is only affected by small variations in predicate selectivity for each query workload. As before, ACORN-1 approaches ACORN-$\gamma$'s search performance.

\begin{figure}[tb]
  \vspace{-.3cm}
  \centering
  \includegraphics[width=0.35\linewidth]{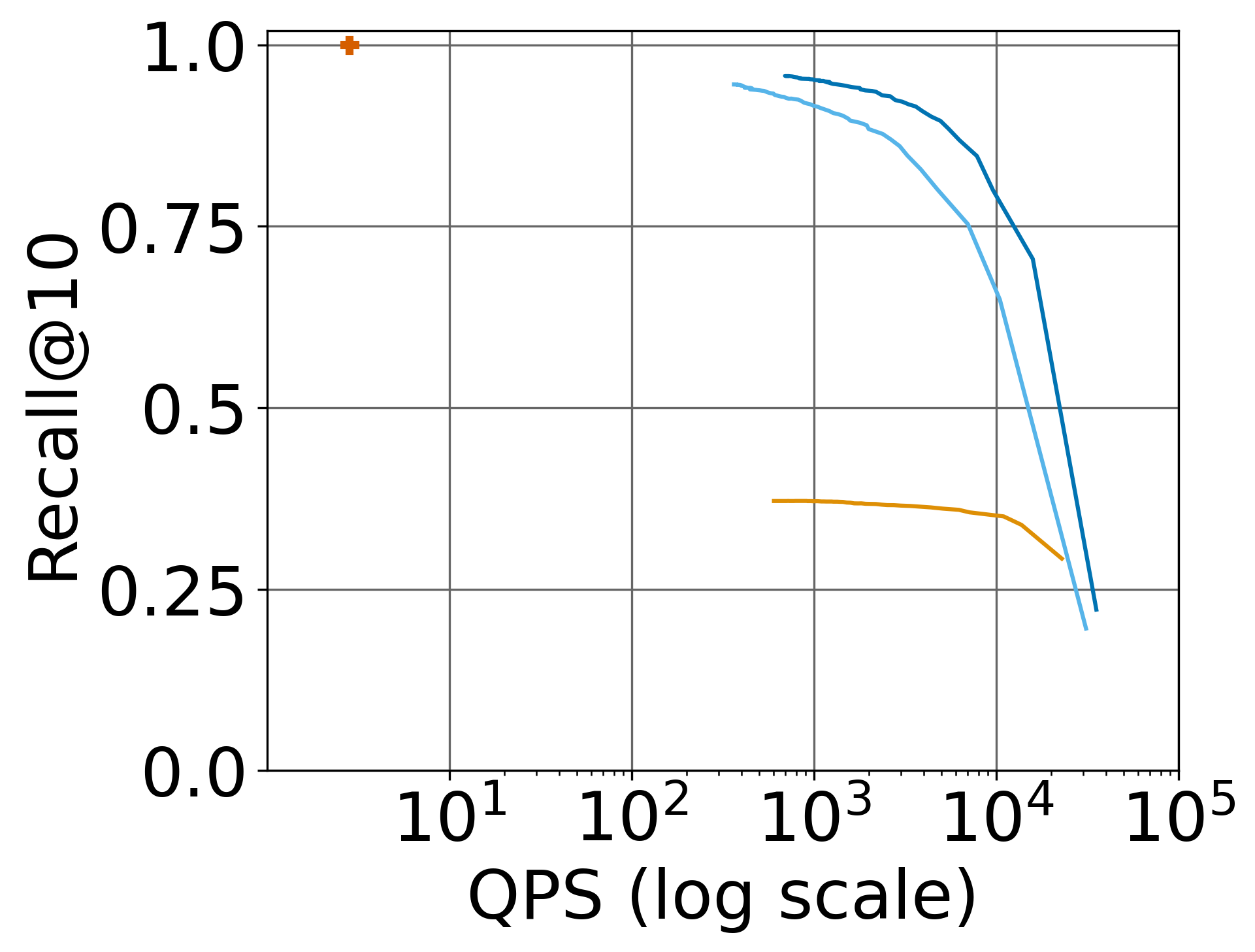}
  \vspace{-.15cm}
  \caption{\small Recall@10 vs QPS on LAION-25M}
  \vspace{0cm}
  \label{fig4:laion_scaling}
\end{figure}

\underline{Scaling Dataset Size:}\label{subsubsec:search_scale}
Figure \ref{fig4:laion_scaling} shows ACORN's search performance on LAION-25M with the no-correlation query workload, demonstrating that the performance gap between ACORN and existing baselines only grows as the dataset size scales. At 0.9 recall, ACORN-$\gamma$ achieves over three orders of magnitude higher QPS than the next best-performing baseline. As before, ACORN-1's search performance approximates that of ACORN-$\gamma$.

\subsection{Index Construction}\label{subsec:construction_oh}
\rev{We will now evaluate ACORN's construction procedure, including its indexing time and space footprint, ACORN-$\gamma$'s compression procedure, and the predicate subgraph quality resulting from ACORN-$\gamma$'s neighbor expansion approach.}

\subsubsection{TTI and Space Footprint}  First, we analyze ACORN's space footprint and indexing time. 
Table \ref{tab:TTI} and \ref{tab:indexsize} show the time-to-index and index size of ACORN-$\gamma$ and ACORN-$1$ compared to the best-performing baselines. \rev{The reported index sizes for each method show the total space footprint of both vector storage and the index itself. All methods are measured using the parameters reported in Section \ref{subsec:benchmarked-methods}.}

We first consider ACORN-$\gamma$'s construction overhead.
Table \ref{tab:TTI} shows that across all datasets, ACORN-$\gamma$'s TTI is at most 11$\times$ higher than HNSW's, and at most 2.15$\times$ higher than that of StitchedVamana, the best performing specialized index.
\rev{Table \ref{tab:indexsize} shows that ACORN-$\gamma$'s index size is at most 1.3$\times$ larger than that of HNSW, and at least $25\%$ smaller than that of StitchedVamana.
The reason for ACORN-$\gamma$'s increased index size and TTI compared to HNSW is it's candidate-edge generation step during construction, which expands each neighbor list. Meanwhile, ACORN-1 achieves the lowest TTI of all listed baselines in table \ref{tab:TTI}, and its index size is at most 1.25$\times$ HNSW's index size and at least $25\%$ smaller than StitchedVamana's index size.} We see that while ACORN-$\gamma$ achieves superior search performance by leveraging a neighbor-list expansion during \emph{construction}, ACORN-1 provides a close approximation at lower TTI and space footprint by instead performing the neighbor-list expansion during \emph{search}. The two algorithms exhibit a trade-off between search performance and construction overhead.

\subsubsection{ACORN-$\gamma$ Pruning}

Given ACORN-$\gamma$'s higher construction overhead, we investigate the efficiency of its predicate-agnostic compression strategy in reducing index construction costs while maintaining search performance.
\rev{First, Table \ref{tab:avg_out_deg} shows ACORN-$\gamma$'s average out-degree per level for each dataset, confirming that compression on level 0 leads to significantly smaller neighbor lists, compared to level without compression, which may have neighbor lists as large as $M\cdot \gamma$.

}

\input{tables/tti}
\input{tables/index_size}
\input{tables/out_degree}

Turning our attention to Figure \ref{fig:pruning_comparison}, we evaluate three different pruning strategies applied to ACORN-$\gamma$'s neighbor lists during construction: 
\textbf{i)} ACORN's predicate-agnostic pruning strategy at varied levels of compression indicated by different $M_{\beta}$ (Mb) values, where $Mb=768$ represents no pruning, and lower values represent more aggressive pruning, 
\textbf{ii)} a metadata-aware RNG-based pruning approach, which is employed by FilteredDiskANN's algorithms, and 
\textbf{iii)} HNSW's metadata-blind pruning.
\rev{We consider TTI, space footprint, the number of candidates edges pruned per node and search performance. 
The figure represents space footprint measured by the average out degree of nodes on on level 0, the level on which each pruning strategy is applied.} In addition, the figure shows search performance measured by recall at 20,000 QPS. We note that the recall ranges of the recall-QPS curve generated by different pruning methods varied significantly, leading us to choose a QPS threshold rather than a recall threshold.
\rev{Interestingly, Figure \ref{fig:pruning_comparison} shows that ACORN's pruning can significantly reduce both the TTI and space footprint by aggressively pruning candidate edges, while maintaining search performance. In comparison, applying HNSW pruning to the index results in significantly degraded hybrid search performance. }Meanwhile the metadata-aware RNG-base pruning results in similar search performance to ACORN-$\gamma$'s pruning, but it is less efficient by TTI and space footprint than ACORN's pruning for small values of $M_\beta$ (e.g., $M_\beta = 32, 64)$. 

\begin{small} 
\begin{figure}[t]
  \centering
  \includegraphics[trim={0.0in 1.800in 0.0in 0.0in}, clip, width=\linewidth]{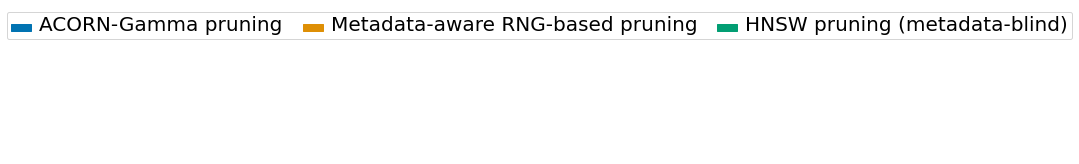}
  \label{fig:pruning_legend}
  \begin{subfigure}[t]{0.33\linewidth}
    \centering
    \includegraphics[width=\linewidth]{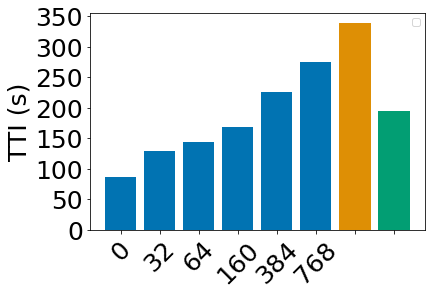}
    \vspace{-.5cm}
    \caption{TTI}
    \vspace{-.1cm}
    \label{fig:pruning_tti}
  \end{subfigure}
  \begin{subfigure}[t]{0.33\linewidth}
    \centering
    \includegraphics[width=\linewidth]{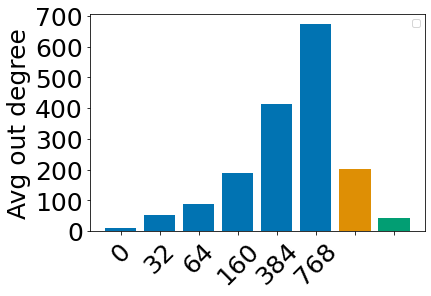}
    \vspace{-.5cm}
    \caption{\rev{Space footprint}}
    \vspace{-.1cm}
    \label{fig:pruning_index_size}
  \end{subfigure}
  \begin{subfigure}[t]{0.33\linewidth}
    \centering
    \includegraphics[width=\linewidth]{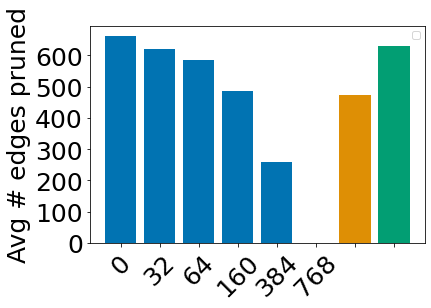}
    \vspace{-.5cm}
    \caption{\rev{\# Edges Pruned}}
    \vspace{-.1cm}
    \label{fig:pruning_num_pruned}
  \end{subfigure}
  \begin{subfigure}[t]{0.33\linewidth}
    \centering
    \includegraphics[width=\linewidth]{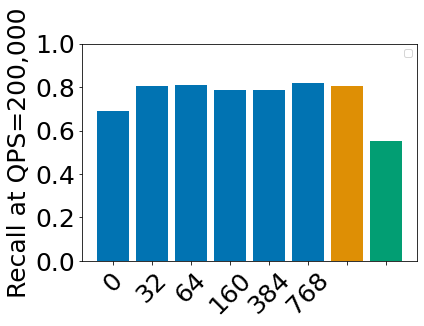}
    \vspace{-.5cm}
    \caption{Search Perf.}
    \vspace{-.1cm}
    \label{fig:acorn_pruning_search}
  \end{subfigure}
  \vspace{-.3cm}
  \caption{\small Comparison of pruning methods on SIFT1M and their impact on TTI (a), space footprint of the index (b), the number of candidate edges pruned (c) and search performance (d). $M_\beta$ values used for ACORN-$\gamma$ are shown along the x-axis.}
  \vspace{-.4cm}
  \label{fig:pruning_comparison}
\end{figure}
\end{small}

\begin{figure}[t]
  \centering
  \includegraphics[trim={0.0in 1.800in 0.0in 0.0in}, clip, width=.8\linewidth]{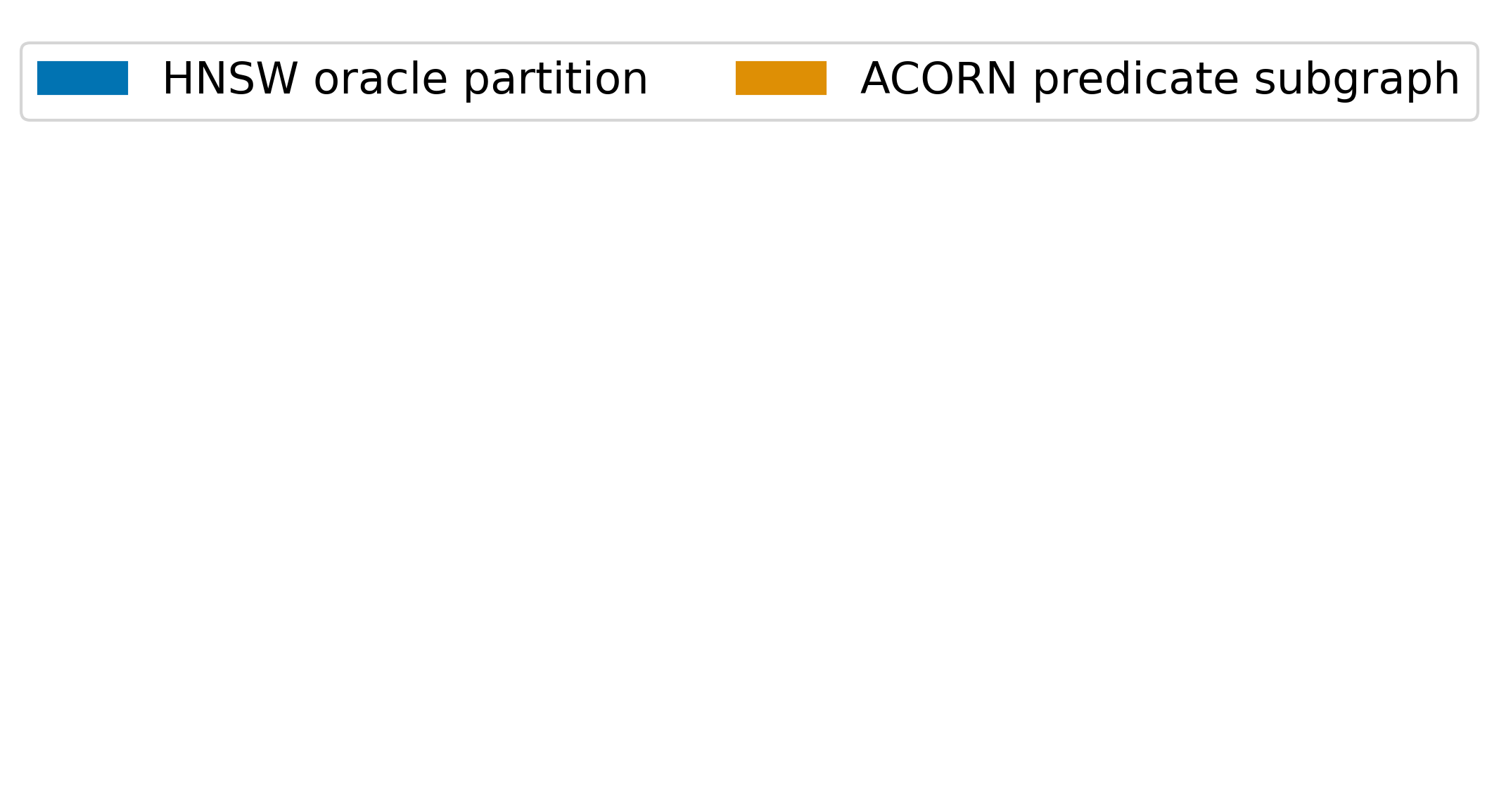}
  \label{fig:graph_quality_legend}
  \vspace{-1.4cm}
  \begin{subfigure}[t]{0.32\linewidth}
    \centering
    \includegraphics[width=\linewidth]{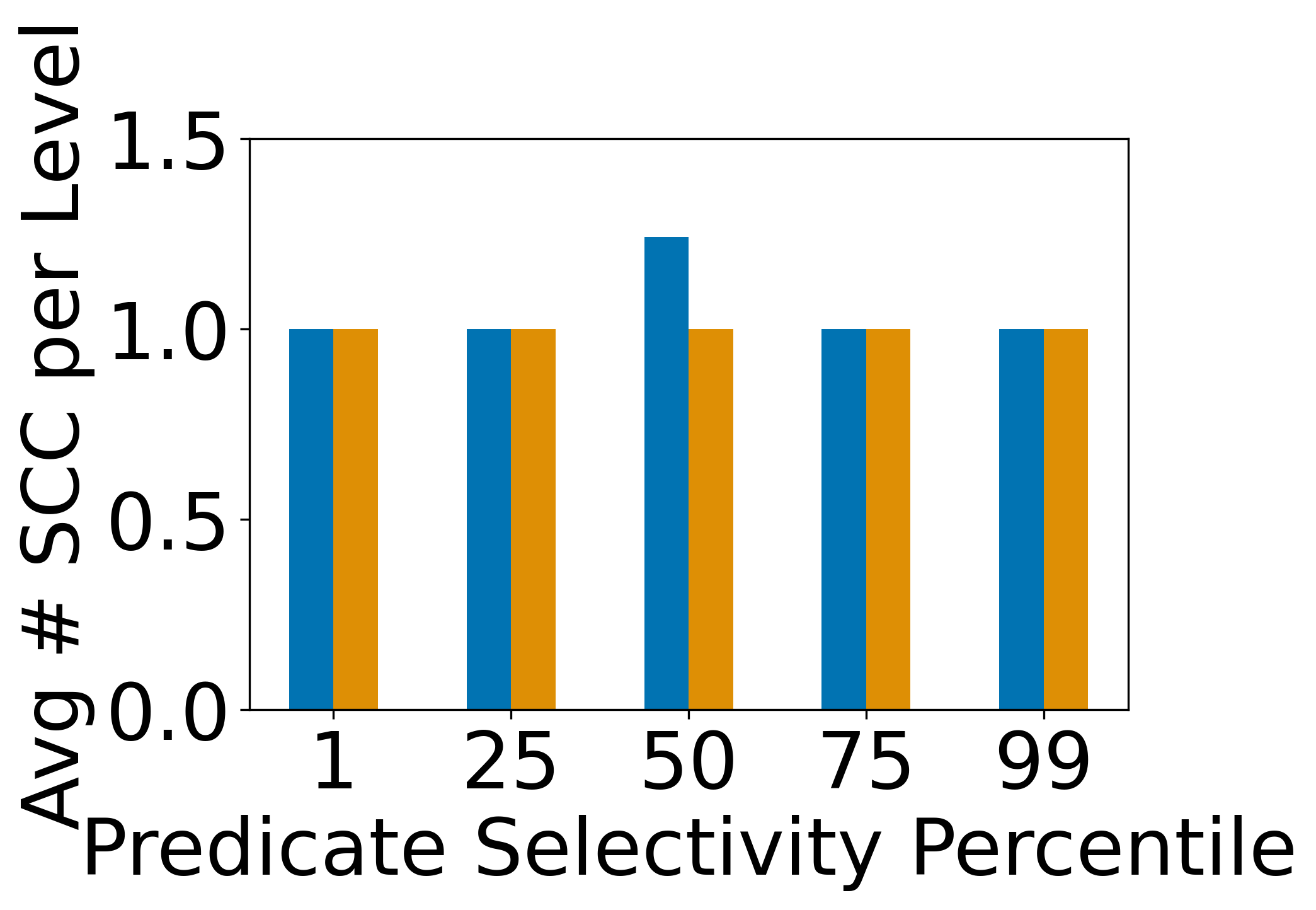}
    \caption{\# SCC}
    \label{fig:num_scc}
  \end{subfigure}
  \begin{subfigure}[t]{0.32\linewidth}
    \centering
    \includegraphics[width=\linewidth]{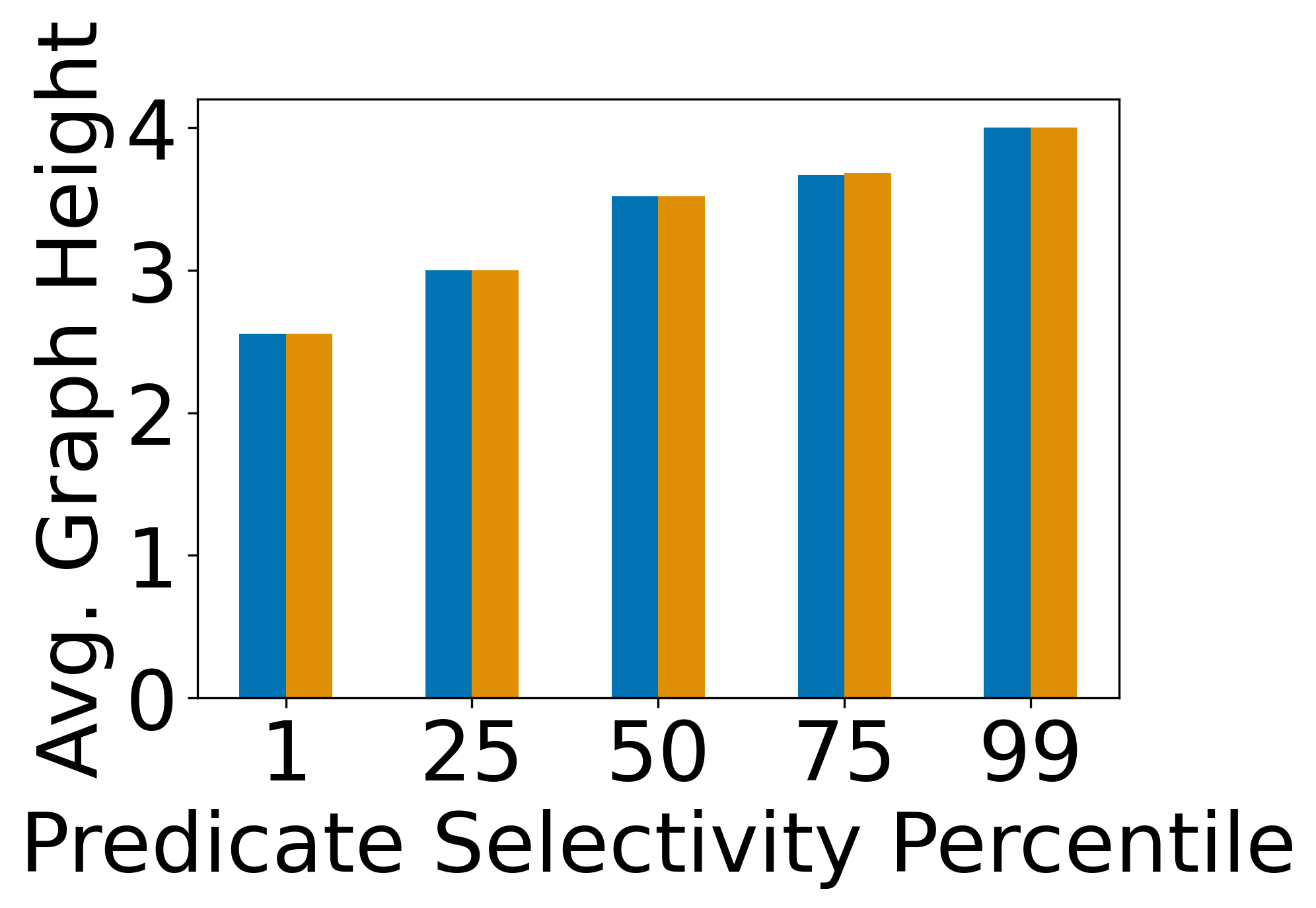}
    \caption{Graph Height}
    \label{fig:graph_height}
  \end{subfigure}
  \begin{subfigure}[t]{0.32\linewidth}
    \centering
    \includegraphics[width=\linewidth]{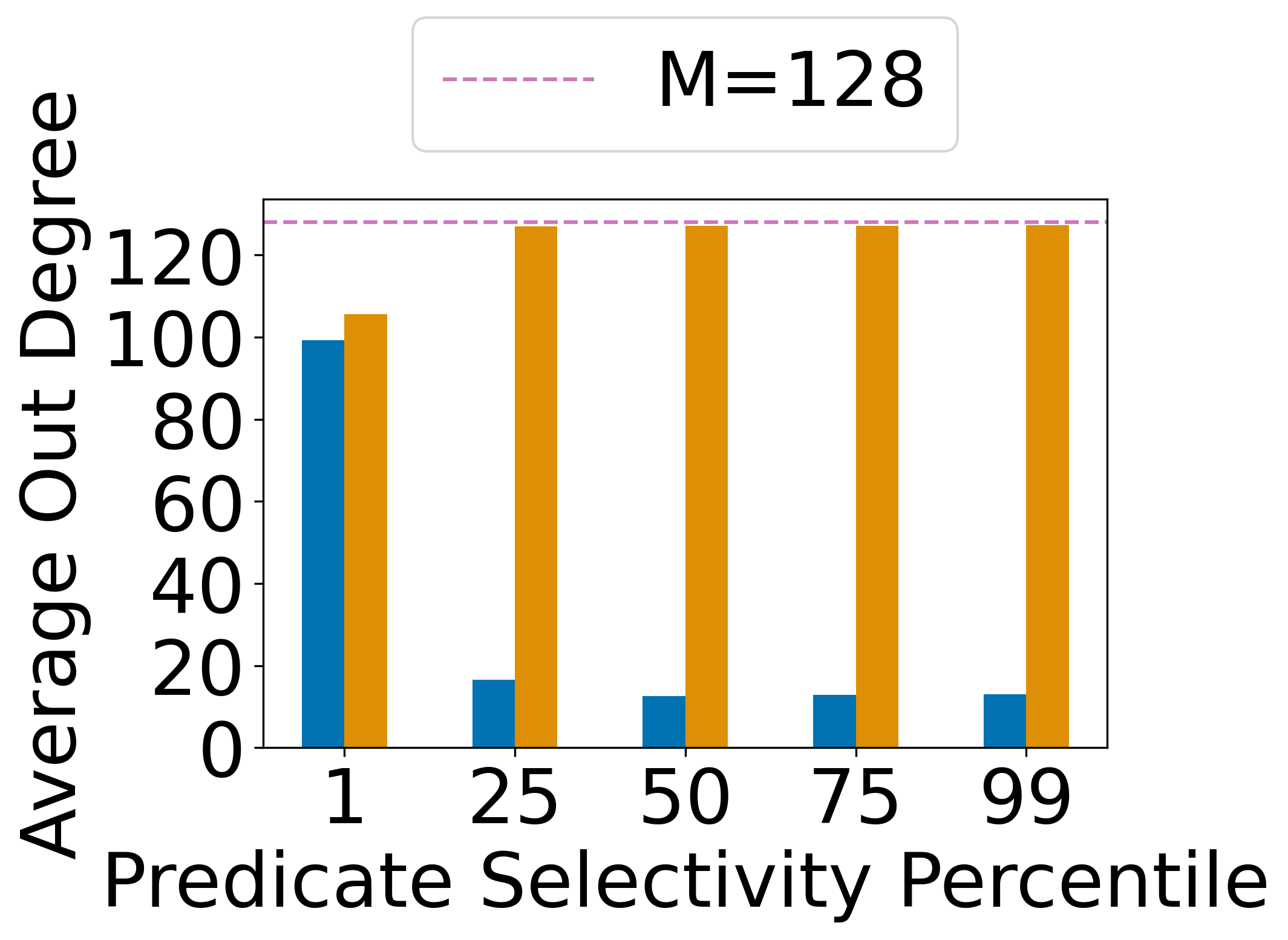}
    \caption{Avg Out Degree}
    \label{fig:out_degree}
  \end{subfigure}
  \vspace{-.4cm}

  \caption{\small \rev{Graph quality of ACORN-$\gamma$ predicate subgraph evaluated by (a) average number of strongly connected components per level, (b) graph height, and (c) average out degree of nodes across uncompressed levels. Results are shown for the TripClick dataset with 1, 25, 50, 75, and 99 percentile selectivity predicates to generate the predicate subgraph and HNSW oracle partition.}}
  \vspace{-.5cm}
  \label{fig:graph_quality}
\end{figure}

\rev{\subsubsection{Graph Quality} Finally, we investigate the graph quality of ACORN-$\gamma$'s predicate subgraphs. 
Figure \ref{fig:graph_quality} compares graph connectivity, graph height, and out degrees for HNSW oracle partitions and ACORN-$\gamma$ predicate subgraphs across varied predicate selectivities on the TripClick dataset's real hybrid search queries. 

From Figure \ref{fig:num_scc}, we see ACORN-$\gamma$'s predicate-subgraph connectivity empirically matches or exceeds that of the HNSW oracle partition across selectivities, demonstrating the effectiveness of ACORN-$\gamma$'s neighbor expansion strategy. 
Next, Figure \ref{fig:graph_height} shows that the controlled hierarchy of ACORN-$\gamma$'s predicate subgraphs emulate that of the HNSW oracle partitions. 
Malkov et al. show that HNSW search performance is sensitive to graph height \cite{malkov_efficient_2018}; thus, this result helps explain ACORN-$\gamma$'s ability to emulate the search efficiency of the oracle partition. 
Lastly, Figure \ref{fig:out_degree} examines the average out degree resulting from  performing the search-time filtering, described in Figure \ref{fig:search_diagram}(a), over the ACORN-$\gamma$ index.
We note that sufficiently high, but bounded, out-degrees are important for emulating HNSW's navigability properties, as discussed in Section \ref{subsec:discussion_search}. 
The figure confirms that ACORN's predicate subgraphs have average out-degrees consistently close to and bounded by $M$.
As expected, the HNSW oracle partition has significantly lower average out-degrees than nodes on ACORN-$\gamma$'s uncompressed levels because HNSW applies RNG-based pruning.
We also note, that the ACORN predicate subgraph with 1 percentile selectivity has lower average out degrees than the other predicate subgraphs because the low selectivity predicates result in fewer than $128$ nodes on the largest uncompressed levels, thus capping the maximum out degree per node below $M=128$. 
Overall, we observe that ACORN-$\gamma$ produces high quality predicate subgraphs, which empirically emulate several HNSW properties related to search efficiency.
}

%% file: tables/datasets.tex
\begin{small}
\begin{table*}[!ht]

  \caption{Datasets}
  \vspace{-.2cm}
  \newcolumntype{C}[1]{>{\centering\arraybackslash}p{#1}}
  \begin{tabular}{c C{1.5cm}C{0.7cm}C{1.6cm}C{2.2cm} C{3.0cm}C{1.7cm}C{1.6cm}}
    \toprule
    \multirow{2}{*}{} & \multicolumn{4}{c}{Base Data} & \multicolumn{3}{c}{Query Workload} \\
    \cmidrule(r){2-5}
    \cmidrule(l){6-8}
    & \small \# Vectors & \small Vector Dim & \small Vector Source Data & \small Structured Data & \small Predicate Operators &  \small Avg. Query Selectivity & \small Predicate Cardinality \\
    \midrule
    \small SIFT1M & \small 1,000,000 & \small 128 & \small images & \small random int. & \small \texttt{equals}($y$) & \small 0.083 & \small 12 \\
    \small Paper & \small 2,029,997 & \small 200 & \small passages & \small random int. & \small \texttt{equals}($y$) & \small 0.083 & \small 12 \\
     \small TripClick & \small 1,055,976 & \small 768 & \small passages & \small clinical area list \& publication date & \small \texttt{contains}($y_1 \vee y_2 \vee ...$) \& \texttt{between($y_1, y_2$)} & \small 0.17, 0.36 \footnotemark & \small $> 10^{8}$ \\
    \small LAION (1M) & \small 1,000,448 & \small 512 & \small images & \small text captions \& keyword list  & \small \texttt{regex-match}($y$) \& \texttt{contains}($y_1 \vee y_2 \vee ...$) &  \small 0.056 - 0.13 \footnotemark & \small $> 10^{11}$ \\
    \small LAION (25M) &  \small 24,653,427 & \small 512 & \small same as above & \small same as above & \small same as above &  \small same as above & \small same as above \\
    \bottomrule
  \end{tabular}
  \label{tab:dataset}

\end{table*}
\end{small}

%% file: tables/distance_comps.tex
\begin{small}

\begin{table}[b] 
    \begin{threeparttable}
        
    \centering
    \vspace{-.2cm}
    \captionsetup{justification=centering}
    \caption{\# Distance Computations to Achieve $0.8$ Recall}
    \vspace{-.20cm}
    \begin{tabular}{ c c c }
    \toprule
     & SIFT 1M & Paper \\
    \midrule
    \small Oracle Partition & \small 398.0  & \small 281.1  \\
    \small ACORN-$\gamma$ & \small 611.0 (+53.5\%) & \small 383.7 (+36.6\%) \\
    \small ACORN-1 & \small 999.6 (+151.0\%) & \small 567.8 (+101.2\%) \\
    \small HNSW Post-filter & \small 1837.8 (+362.6\%) & \small 1425.5 (+406.2\%) \\
    \bottomrule
    \end{tabular}
    \begin{tablenotes}
        \Small \item * Percentage difference is shown in parenthesis and is relative to oracle partition method
    \end{tablenotes}
    \label{tab:dist_comps}
    \end{threeparttable}

\end{table}

\end{small}



%% file: tables/tti.tex
\begin{small}
    
\begin{table}[!t]
\vspace{-0.4cm}
\caption{TTI (s)}
\vspace{-0.2cm}
\resizebox{0.9\columnwidth}{!}{%
  \begin{tabular}{cccccc}
    \toprule
     & TripClick & LAION-1M & LAION-25M & Sift1M & Paper\\
    \midrule
    ACORN-$\gamma$ & 9902.9 & 835.8 & 38,007.5 & 148.9 & 255.6 \\
    ACORN-1 & 322.9 & 25.9 & 705.3 & 8.6 & 27.0 \\
    HNSW & 891.0 & 32.9 & 1,147.2 & 11.3 & 29.2 \\
    FilteredVamana & NA & NA & NA & 18.3 & 51.9 \\
    StitchedVamana & NA & NA & NA & 69.2 & 189.7 \\
    \bottomrule
  \end{tabular}%
  }
  \vspace{-0.2cm}
  \label{tab:TTI}
\end{table}

\end{small}

%% file: tables/index_size.tex
\begin{small}
    
\begin{table}[!t]
\vspace{-0.1cm}
\caption{Index Size (GB)}
\vspace{-0.2cm}
  	\resizebox{0.9\columnwidth}{!}{%
  \begin{tabular}{cccccc}
    \toprule
     & TripClick & LAION-1M & LAION-25M & Sift1M & Paper\\
    \midrule
    ACORN-$\gamma$ & 4.9 & 2.4  & 59  & 0.98  & 2.5  \\
    ACORN-1 & 4.6 & 2.3 & 59 & 0.93 & 2.4 \\
    HNSW & 4.1  & 2.2  & 54  & .75 & 2.1  \\
    Flat Index & 3.1  & 1.9  & 47  & .51 & 1.6  \\
    FilteredVamana & NA & NA & NA & .61  & 1.8 \\
    StitchedVamana & NA & NA & NA & 1.3  & 3.5  \\
    \bottomrule
  \end{tabular}%
  }
  \vspace{-0.4cm}
  \label{tab:indexsize}
\end{table}
\end{small}

%% file: tables/out_degree.tex
\begin{small}
    
\begin{table}[!t]
\caption{ACORN-$\gamma$ Average Out Degree}
\vspace{-0.2cm}
  	\resizebox{0.9\columnwidth}{!}{%
  \begin{tabular}{cccccc}
    \toprule
     & TripClick & LAION-1M & LAION-25M & Sift1M & Paper\\
    \midrule
    Level 0 (compressed) & 191 & 50.1 & 49.4 & 87.5 & 86.0\\
    Level 1 & 8,075 & 960 & 960 & 384 & 384 \\
    Level 2 & 54.0 & 919 & 937 & 363 & 359 \\
    Level 3 & 0 & 25.3 & 689 & 25.3 & 57.4 \\
    Level 4 & NA & 0 & 16 & 0 & 1.0 \\
    $M\cdot \gamma$ & 10,240 & 960 & 960 & 384 & 384 \\
    $M_\beta$ & 128 & 32 & 32 & 64 & 64 \\
    \bottomrule
  \end{tabular}%
  }
  \vspace{-0.3cm}
  \label{tab:avg_out_deg}
\end{table}

\end{small}

%% file: Sections/RelatedWork.tex
\section{Related Work}

\emph{\textbf{Pre- \& Post-filtering-based Systems.}} Many hybrid search systems rely on pre- and post-filtering. While several systems have developed pre-processing methods to perform \emph{faster filtering} during search, these systems fail to reduce the \emph{excessive and expensive distance computations} which bottleneck performance.
Weaviate \cite{noauthor_filtered_nodate} creates an inverted index for structured data ahead of time, then uses it at query time to create a bitmap of eligible candidates during post-filtering.
Milvus \cite{wangmilvus2021} likewise creates an approved list of points by maintaining a distribution of attributes over the dataset in order to map commonly used query filters to a list of approved points before performing pre- or post-filtering.
Several space-partitioning indices like FAISS-IVF \cite{baranchuk_revisiting_2018, johnson_billion-scale_2017} and LSH \cite{andoni_practical_2015} store metadata information in the index, allowing them to rapidly filter entities during post-filtering.
Despite the optimized filtering steps in each of these approaches, the core problems of pre- and post-filtering remain, particularly for low correlation or selectivity predicates.

\emph{\textbf{Specialized Indices.}} Alternatively, several recent works develop novel graph-based algorithms for hybrid search, often improving performance for a constrained set of predicates.
NHQ \cite{wang_navigable_2022} encodes attributes alongside vectors, and then uses a "fusion distance" during search that accounts for vector distances as well as attribute matches. This approach supports only equality query predicates and assumes each dataset entity has only one structured attribute. 
Filtered-DiskANN \cite{gollapudi_filtered-diskann_2023} proposes two algorithms: FilteredVamana and StitchedVamana. Both methods constrain the query filter cardinality to about $1,000$ with only equality predicates so that the index construction steps can use this knowledge to appropriately generate and prune candidate edge lists.
Similarly HQI \cite{mohoney_high-throughput_2023} optimizes batch query-processing by assuming a limited cardinality of $20$ query predicates to design an efficient partitioning scheme.  
\rev{On the otherhand, Qdrant \cite{vasnetsov_filtrable_nodate} proposes to densify an HNSW graph and perform a filtered greedy search. While this approach aligns intuitively with ACORN's neighbor-list expansions during construction, Qdrant's proposal inadvertently flattens the graph by directly increasing the HNSW parameter $M$, which impacts HNSW's level normalization constant. Malkov et al. show that HNSW's performance is sensitive to its number of levels, and flattening the graph degrades search performance \cite{malkov_efficient_2018}. In addition, Qdrant's proposed method does not provide a solution for dealing with the increased memory overhead after creating a denser HNSW. }

%% file: Sections/Conclusion.tex
\vspace{-.2cm}
\section{Conclusion}
We proposed ACORN, the first approach for efficient hybrid search across vectors and structured data that supports large and diverse sets of query predicates. ACORN uses a simple, yet effective, search strategy based on the core idea of \emph{predicate subgraph traversal}. We presented two indices, ACORN-$\gamma$ and ACORN-1, that implement this search strategy by modifying the HNSW indexing algorithm.
Our results show that ACORN achieves state-of-the-art hybrid search performance on both prior benchmarks, involving simple, low-cardinality query predicate sets, as well as more complex benchmarks involving new predicate operators and high cardinality predicate sets. 
Across both types of benchmarks, ACORN-$\gamma$ achieves 2--1,000$\times$ higher QPS at 0.9 recall than prior methods, and ACORN-1 approximates ACORN-$\gamma$'s search performance with 9--53$\times$ lower TTI for resource-constrained settings.